\documentclass{aa} % for a referee version
\usepackage{graphicx}
\usepackage{url}

\usepackage{color}

\bibpunct{(}{)}{;}{a}{}{,} 
\usepackage{psfrag}
\usepackage{version}

\usepackage{txfonts}
\usepackage{float}
\usepackage[colorlinks=true,linkcolor=blue,citecolor=blue,urlcolor=black]{hyperref}
\usepackage[normalem]{ulem}

\bibpunct{(}{)}{;}{a}{}{,} 

\usepackage{tabularx}
\usepackage{tablestyles}
\usepackage[detect-all=true]{siunitx}
\DeclareSIUnit\astronomicalunit{au}
\DeclareSIUnit\parsec{pc}
\usepackage{natbib,twoopt}

\begin{document} 

	\title{Spatial distribution of the aromatic and aliphatic carbonaceous nano-grain features in the protoplanetary disk around HD~100546}
	\author{E. Habart\inst{1},
	      T. Boutéraon\inst{1},
	      R. Brauer \inst{1,2},
          N. Ysard\inst{1},
          E. Pantin\inst{2},
          A. Marchal\inst{1,3},
          A.P. Jones\inst{1}
          }

	\institute{\inst{1}Universit\'e Paris-Saclay, CNRS,  Institut d'Astrophysique Spatiale, 91405, Orsay, France. \\
	 \email{emilie.habart@ias.u-psud.fr} \\
               \inst{2}IRFU/SAp Service D'Astrophysique, CEA, Gif-sur-Yvette, France \\
                \inst{3}Canadian Institute for Theoretical Astrophysics, University of Toronto, 60 St. George Street, Toronto, ON M5S 3H8, Canada\\}

	\date{Accepted by A\&A, August 25, 2020}

  \abstract
   {Carbonaceous nano-grains are present at the surface of protoplanetary disks around Herbig Ae/Be stars, where most of the ultraviolet energy from the central star is dissipated. Efficiently coupled to the gas, they are unavoidable in the understanding of the physics and chemistry of these disks. Furthermore, nano-grains are able to trace the outer flaring parts of the disk, and possibly the gaps from which the larger grains are missing. However, their evolution through the disks, from internal to external regions are not yet well understood. 
   }
   {Our aim is to examine the spatial distribution and evolution of the nano-dust emission in the emblematic (pre-)transitional protoplanetary disk HD~100546. This disk shows many structures (annular gaps, rings, and spirals) and reveals very rich carbon nano-dust spectroscopic signatures (aromatic, aliphatic) in a wide spatial range of the disk ($\sim 20-200$~au).}
   {We analyse adaptive optics spectroscopic observations in the 3-4~$\mu$m range (angular resolution of $\sim 0.1$"), as well as, imaging and spectroscopic observations in the 8-12~$\mu$m range (angular resolution of $\sim 0.3$"). The hyper-spectral cube is decomposed into  a sum of spatially coherent dust components using the Gaussian decomposition algorithm {\tt ROHSA} (Regularized Optimization for Hyper-Spectral Analysis). We compare the data to model predictions using The Heterogeneous dust Evolution Model for Interstellar Solids (THEMIS) integrated into the radiative transfer code POLARIS by calculating the thermal and stochastic heating of micro- and nanometre-sized dust grains for a given disk structure.}
   {We find that the aromatic features at 3.3, 8.6, and 11.3~$\mu$m, as well as, the aliphatic features between 3.4 and 3.5~$\mu$m are spatially extended, with each band showing a specific morphology dependent on the local physical conditions. The aliphatic-to-aromatic band ratio, 3.4/3.3 increases with the distance from the star from $\sim 0.2$ (at 0.2" or 20~au) to $\sim 0.45$ (at 1" or 100~au) suggesting UV processing. 
   In the 8-12~$\mu$m observed spectra, several features characteristic of aromatic particles and crystalline silicates are detected with their relative contribution changing with the distance to the star.
   The model predicts that the features and adjacent continuum are due to different combinations of grain sub-populations, with in most cases a high dependence on the intensity of the UV field. Shorter wavelength features are dominated by the smallest grains ($a < 0.7$~nm) throughout the disk, while at longer wavelengths what dominates the emission close to the star is a mix of several grain populations, and far away from the star is the largest nano-grain population.
   }
   {With our approach that combines both, observational data in the near- to mid-IR and disk modelling, we deliver constraints on the spatial distribution of nano-dust particles as a function of the disk structure and radiation field.}

\authorrunning{}
\titlerunning{IR band features around HD~100546}
\maketitle

%--------------------------------------------------------------------
\section{Introduction \label{sec_intro}}

The objective of this article is to study the spatial distribution and possible changes in the properties of carbon nano-dust in protoplanetary disks (PPDs). Carbon nano-dust, detected under more or less organised structures and different ionisation states, constitutes a major component of dust in the interstellar and circumstellar environments.
Vibrational emission bands in the near- to mid-IR from nano-carbon dust have been observed towards PPDs around most of the Herbig Ae stars, about half of the Herbig Be stars, and a few T-Tauri stars \citep[e.g.][]{Brooke1993,acke_iso_2004, Acke2010, Seok2017}.
Contrary to large grains, these tiny and numerous carbon grains are well coupled to the gas and do not settle towards disk midplanes. It results in different spatial distributions with tiny grains present at the disk surfaces \citep[e.g.][]{Meeus2001, Habart2004, Lagage2006} and in the cavity or gaps from which the pebbles are missing \citep[e.g.][]{Geers2007, Kraus2013,Klarmann2017, Kluska2018, Maaskant2013}.
The presence of the very small carbon grains in the irradiated disk layers may have strong consequences  \citep[e.g.][]{gorti2008}. As in the irradiated regions of the interstellar medium, they are the prevalent contributors to the energetic balance because they are very efficient at absorbing UV photons and heating the gas via the photoelectric effect. The highest fluxes of lines tracing the warm gas (e.g., [OI] 63 and [OI] 145~$\mu$m, H$_2$ 0-0 S(1), high-J CO) are found in PPDs which show a large amount of flaring and high aromatic band strength \citep[e.g.][]{Meeus2013}. 
Moreover, due to their large effective surface area, they may dominate the catalytic formation of key molecules as H$_2$, as well as the charge balance. The disk structure further may depend on the level of nano-grains coupling with the gas. 
Characterising the size and properties of these tiny grains through the disks, from internal to external regions,
is thus of prime importance to understand the structure and evolution of PPDs.

In a recent paper, \citet{Bouteraon2019} showed the presence of several spatially extended near-IR spectral features that are related to aromatic and aliphatic hydrocarbon material in PPDs around Herbig stars, from 10 to 50-100~au, and even in inner gaps that are devoid of large grains. The correlation between aliphatic and aromatic CH stretching bands suggested common carriers for all features. Since these hydrocarbon nano-particles are a priori easily destroyed by UV photons \citep[e.g.][]{Munoz2001, Mennella2001, Gadallah2012}, this probably implies that they are continuously replenished at the disk surfaces. In the continuity of the \citet{Bouteraon2019} study, we investigate here the nano-dust properties in disks thanks to a  decomposition tool applied to spectra, complementary observations at longer mid-IR wavelengths, and the coupling of THEMIS \citep[The Heterogeneous dust Evolution Model for Interstellar Solids, ][]{Jones2017} with the radiative transfer code POLARIS \citep[POLArized RadIation Simulator, ][]{Reissl2016, brauer_magnetic_2017-1}. 
We focus on the emblematic protoplanetary disk HD~100546, in transitional phase from a gas-rich to a dust debris disk \citep[e.g.][]{Bouwman2003}.

The paper is organised as follows.
Section~\ref{sec_HD100546} provides a description of the protoplanetary disk HD~100546. 
In Sect.~\ref{sec_obs}, we analyse the adaptive optics spectroscopic observations in the L band obtained with NAOS-CONICA (NaCo) at VLT (Very Large Telescope) using {\tt ROHSA} for the decomposition. Imaging and spectroscopic observations obtained with VISIR (VLT Imager and Spectrometer for mid Infrared) at VLT are also presented and analysed.
In Sect.~\ref{sec_polaris}, the disk modelling with the THEMIS dust model and the radiative transfer code POLARIS are presented.
In Sect.~\ref{sec_model_obs}, we compare the model predictions to the NaCo and VISIR observations.
In Sect.~\ref{sec_discussion}, our results concerning the nano-dust evolution are discussed.

\begin{table*}
\caption{Band centre ($\lambda_0$, $\nu_0$) and FWHM variations in laboratory experiments (see references in \citet{Jones2013} and \citet{Bouteraon2019}).}             
\label{tab_versat}      
\centering        
\begin{tabular}{c c c c c}     
\hline \hline 
 band & $\lambda_{0}$ [$\mu$m]  & FWHM [$\mu$m] & $\nu_0$ [cm$^{-1}$] & FWHM [cm$^{-1}$]\\
\hline
 sp$^2$ CH aro & 3.268 - 3.295  & 0.005 - 0.057 & 3060 - 3035 & 5 - 53.1 \\
 sp$^2$ CC aro & 8.60 & 0.665 & 1163 & 90 \\
 sp$^2$ CH aro & 11.236 - 11.364 & 0.252 - 0.517 & 890 - 880 & 20 - 40 \\
 \hline
sp$^2$ CH$_2$ ole. asy str & 3.240 - 3.249 & 0.014 - 0.045 & 3089 - 3078 &  13.3 - 42.4\\
 sp$^2$ CH ole. & 3.311 - 3.344 &  0.008 - 0.062 & 2990 - 3020  & 7 - 56.3  \\
 sp$^2$ CH$_2$ ole. sym str & 3.350 - 3.396  & 0.003 - 0.028 & 2985 - 2945 & 3 - 25 \\
 sp$^2$ CH$_3$ ali. asy str & 3.378 - 3.384 & 0.008 - 0.033 & 2960 - 2955  & 7 - 29.3 \\
 sp$^3$ CH$_2$ ali. asy str & 3.413 - 3.425 & 0.006 - 0.034 & 2930 -2920  & 5 - 28.9 \\
 sp$^3$ CH tertiarty ali. / Fermi resonance & 3.425 -3.47 & 0.006 - 0.036 & 2920 - 2882 & 5 - 30 \\
 sp$^3$ CH$_3$ ali. sym str & 3.466 - 3.486 & 0.006 - 0.034 & 2885 - 2869 & 5 - 27.8 \\
 sp$^3$ CH$_2$ ali. sym str & 3.503 - 3.509 & 0.006 - 0.051 & 2855 - 2850 & 5 - 41.8 \\
 sp$^3$ CH$_2$ ali. sym str wing & 3.552  & 0.039 - 0.068 & 2815 & 31 - 53.6\\
\hline                  
\end{tabular}
\end{table*}

\section{HD~100546} \label{sec_HD100546}

HD~100546 is one of the closest very well studied Herbig Be stars \citep[$d =110 \pm 4$~pc,][]{Gaia2018}. Its shows clear evidence of a large flared disk and, based on Hubble Space Telescope (HST) and ground-based high-contrast images, an elliptical structure was detected that extends up to 350–380~au \citep{Augereau2001}. Furthermore, multiple-armed spiral patterns were identified as well \citep{Grady2001, Ardila2007, Boccaletti2013, Avenhaus2014}. 
The disk position angle measurements found in the literature range from $\sim 130$ to 160$^{\circ}$ \citep{Grady2001, Pantin2000, Ardila2007, Panic2014, Augereau2001, Avenhaus2014} and the inclination of the disk is smaller than 50$^{\circ}$ \citep{Avenhaus2014}. 
An inner dust disk extending from $\sim 0.2$ to $\sim 1-4$~au was resolved using near-IR interferometry \citep{Benisty2010, Tatulli2011, Mulders2013, Panic2014}. The (pre-)transitional nature of HD~100546 was initially proposed by \citet{Bouwman2003} based on a spectral energy distribution (SED) analysis. The presence of a gap extending up to 10-15~au has been confirmed by mid-IR interferometry \citep{Liu2003, Panic2014}, spectroscopy in the UV and near-IR \citep{Grady2005, Brittain2009, VanDerPlas2009}, and high-resolution polarimetric imaging in the optical and near-IR \citep{Avenhaus2014, Quanz2015, Garufi2016, Follette2017}. Recent ALMA observations reveal an asymmetric ring between $\sim 20-40$~au with largely optically thin dust emission \citep{Pineda2019}. A central compact emission is also detected, which arises from the inner central disk, that, given its mass and the accretion rate onto the star, must be replenished with material from the outer disk \citep{Pineda2019}. 
\citet{Miley2019} presented the first detection of C$^{18}$O in this disk, which spatially coincides with the spiral arms, and derived a lower-limit on the total gas mass (around 1\% of the stellar mass) and a gas-to-dust mass ratio in the disk of $\sim 20$ assuming ISM abundances of C$^{18}$O relative to H$_2$. 
HD~100546 is also one of the few cases where protoplanet candidates have been suggested \citep[e.g.][]{Quanz2015, Currie2015} although this is still being debated \citep[e.g.][]{Follette2017, Rameau2017, Perez2019}. Recently, \citet{Perez2019} detected a compact 1.3 mm continuum dust emission source which lies in the middle of the HD~100546 cavity (0.051" from the central star) which is compatible with circumplanetary disk emission.

On the other hand, this disk with high density and temperature coupled with high UV flux on its surface presents very rich IR spectral features and variations in the chemical and physical dust structural properties.
HD~100546 is the first disk in which crystalline silicates and intense aromatic bands between 3 and 13~$\mu$m have been detected \citep{Hu1989, Malfait1998}. The carriers of these bands can be attributed to polyaromatic species and associated with an emission mechanism based on polycyclic aromatic hydrocarbons (PAHs) photophysics. These intense bands are blended with many fainter sub-bands \citep[e.g.][]{Acke2010, Bouteraon2019} and are spatially extended on a few 100~au scale \citep{VanBoekel2004, Habart2006}. 
In \citet{Bouteraon2019}, we presented the diversity of the sub-features in the 3-4~$\mu$m range where C-H vibrational modes are observed.  
These modes are particularly interesting since they characterise the bonds between carbon and hydrogen atoms varying according to their local environment and strongly depending on the hydrocarbon grain sizes. Constraints can thus be derived for the grain size distribution and the hydrogen-to-carbon ratio in the grains, depending on their more or less aromatic or aliphatic nature. 

\citet{Seok2017} extracted the global mid-IR nano-carbon dust spectrum of HD~100546 and analysed it using the \citet{Li2001} and \citet{draine_infrared_2007} astro-PAH model. They found that most of the PAHs are neutral, with a low ionisation fraction of 0.2, and a mass distribution peaked at 5.64~$\AA$ (log-normal size distribution centred at 0.5~$\AA$ with a width of 0.2), similar to what is required to explain diffuse ISM observations. These results are in agreement with the analysis made by \citet{Bouteraon2019} using the THEMIS amorphous hydrogenated carbon nano-grains. Assuming that all PAHs are located at 10~au from the central star, \citet{Seok2017} found a total PAH mass in the disk of $M_{{\text{PAH}}} = 5.37 \times 10^{-6} M_{\oplus}$. 

Finally, HD~100546 is one of the few Herbig Ae/Be stars to show a strong luminosity in the aromatic bands and warm gas lines such as the rotational and rovibrational lines of H$_2$ \citep{Carmona2011}, CH$^+$ \citep{Thi2011} and CO \citep{Meeus2012, Meeus2013} at the same time. This strongly suggests that the carriers of the aromatic bands which efficiently absorb the stellar UV radiation are one, if not the main, source of the gas heating through photoelectric effect in the warm disk surfaces \citep[e.g.][]{Meeus2013}.

\section{NaCo and VISIR observations}
\label{sec_obs}

\subsection{NaCo observations and data decomposition using {\tt ROHSA}}

\subsubsection{Observations}

NaCo observations were performed using a long slit in the L-band, between 3.20 and 3.76 $\mu$m with the adaptive optics system NaCo at the VLT. The on-sky projection of the slit is 28"-long and 0.086"-wide, which corresponds to the diffraction limit in this wavelength range. The pixel scale is 0.0547" and the spectral resolution is $R = \lambda / \Delta \lambda \sim 1000$. We took nine slit positions, one centred on the star and the other slits shifted by a half width. Nine positions allowed to extract a spectral cube on an area star-centred of $2^{\prime\prime}\times0.354^{\prime\prime}$. 
The long slit was aligned with the major axis of the disk as resolved in scattered light \citep{Augereau2001, Grady2001} with a position angle of 160$^{\circ}$ measured north to east. 
The data set reference is 075.C-0624(A) and observations characteristics and data reduction are summarised in \citet{Bouteraon2019}.

\subsubsection{Spectral cube decomposition using {\tt ROHSA}}

The NaCo spectral cube was decomposed using {\tt ROHSA} in order to produce maps in the various dust features observed in the 3.20 to 3.76~$\mu$m  spectral range. {\tt ROHSA} is based on a regularised non-linear criterion that takes into account the spatial coherence of the emission \citep{marchal2019}. To fit the cube, a multi-resolution process from coarse to fine grid is used. First, a Gaussian decomposition with a fix number of Gaussians N is performed on a spatially averaged version of the data (corresponding at the first iteration to the mean spectrum of the cube). So far, the method is similar to the one described in \citet{Bouteraon2019}. The solution is then interpolated step by step at higher resolution until the initial resolution of the observations is reached. At each step of this process, the algorithm converges towards a solution that, for each Gaussian, is spatially coherent. Additionally, {\tt ROHSA} minimises the variance of the dispersion of each component meaning that on average a Gaussian will have a similar width across the disk. It turned out that due to a strong spectral blinding of the signatures, this minimisation is essential for their separation.
This commanded spatial coherence allows to remove the random noise. Note that the central parts of the spectral cube are not provided to {\tt ROHSA} because the signal-to-noise ratio in the bands is too low due to a strong continuum emission.

\subsubsection{Results \label{subsub_rohsa-res}}

\begin{figure*}[htbp]
    \centering
    \includegraphics[width=17cm]{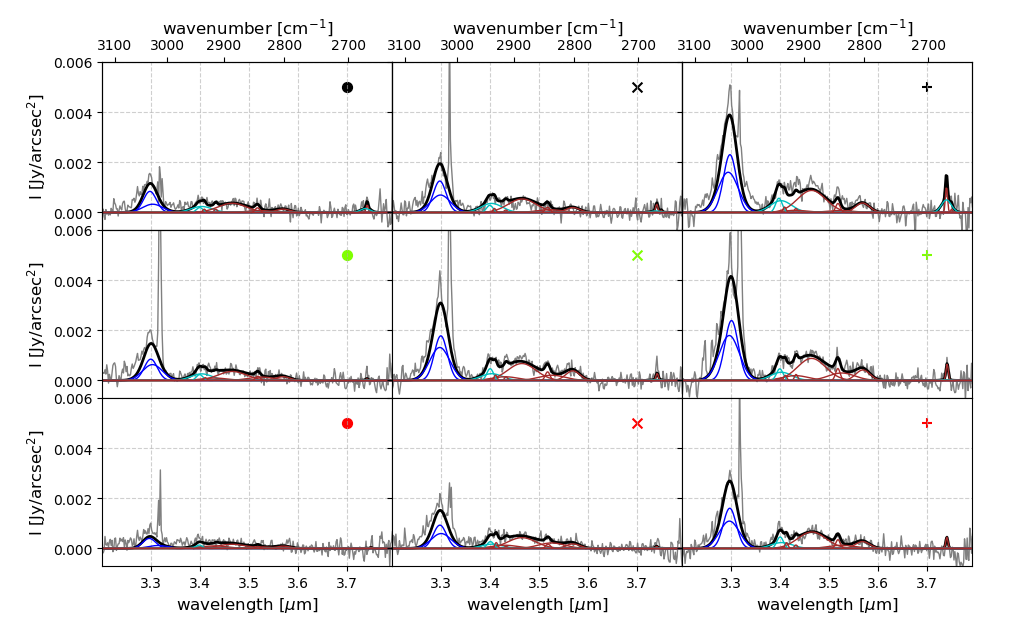}
    \caption{Mosaic of near-IR emission spectra of HD~100546 between 3.2 and 3.8~$\mu$m. Their location (coloured symbols) are reported in Fig.~\ref{fig_rohsa_maps}. The result of the decomposition with {\tt ROHSA} (black) is overplotted on NaCo data (grey). Gaussians related to the 3.3~$\mu$m aromatic feature are in blue, and those related to the 3.4~$\mu$m aliphatic feature in cyan. Other features are plotted in brown.}
    \label{fig_rohsa_mosaic}
\end{figure*}

\begin{figure*}[htbp]
    \centering
    \includegraphics[width=17cm]{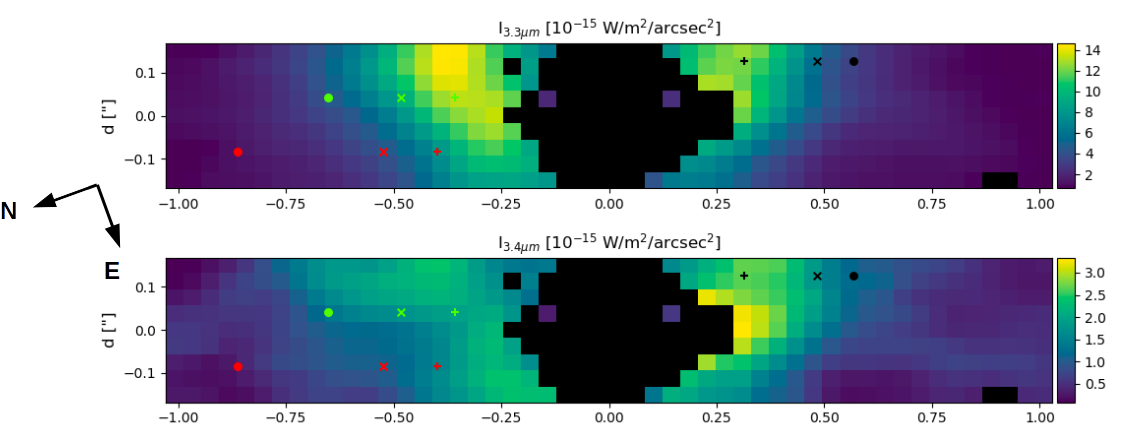}
    \caption{3.3~$\mu$m and 3.4~$\mu$m band emission maps. Intensities are integrals of the Gaussians related to the bands as seen in Fig.~\ref{fig_rohsa_mosaic}. Positions of the spectra in Figure~\ref{fig_rohsa_mosaic} are represented by the coloured symbols.}
    \label{fig_rohsa_maps}
\end{figure*}

Figure~\ref{fig_rohsa_mosaic} shows the result of the decomposition for spectra coming from different locations in the disk (Fig. \ref{fig_rohsa_maps}). 
Twenty-eight Gaussians are needed to fully describe the signal and to fit the spectra towards the entire cube. We assume that a physical signature can be reproduced by several Gaussians. We gather Gaussians according to their central wavelength as in \citet{Bouteraon2019} who considered six features related to carbonaceous materials: one aromatic signature at 3.3~$\mu$m, and five aliphatic ones at 3.4, 3.43, 3.46, 3.52, and 3.56~$\mu$m (Fig. \ref{fig_rohsa_other-features}). The rest of Gaussians correspond to the hydrogen recombination and telluric lines. We move aside the Gaussians expected to be related to telluric bands. 

For the 3.3~$\mu$m aromatic band, we consider two Gaussians, one narrow centred at 3.297~$\mu$m and one broader centred at 3.306~$\mu$m. As discussed in \citet{Bouteraon2019}, these two components could originate from two distinct kinds of bonding: aromatic and olefinic, respectively. For the 3.4~$\mu$m aliphatic band, we consider three Gaussians, two narrow ones centred at 3.398 and 3.409~$\mu$m, and one broader centred at 3.403~$\mu$m. In this study, we focus only on these two features since they are brighter and less blended than the others.

\begin{figure}[htbp]
    \resizebox{\hsize}{!}{\includegraphics[width=0.9\textwidth,height=0.9\textheight,keepaspectratio]{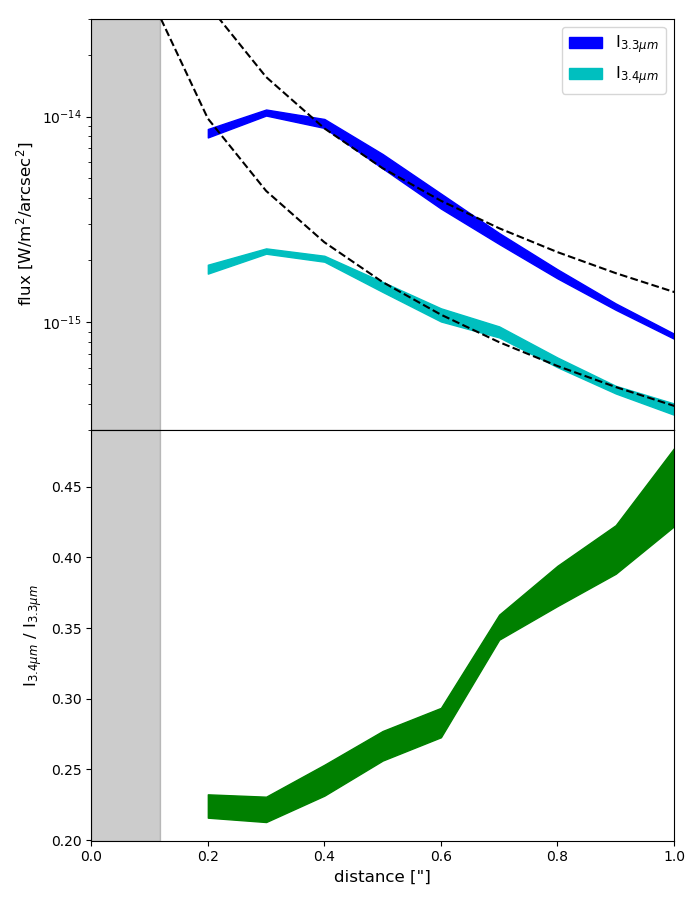}}
    \caption{Top panel: $I_{3.3\mu{\rm m}}$ (blue) and $I_{3.4\mu{\rm m}}$ (cyan) band emission according to the distance to the star. Values are the sum of the integrals of the Gaussians related to each feature and averaged for different distances from the star in 0.1" step. Normalised $1/d^2$ functions are also plotted (black dotted lines) to help the reader. Bottom panel: $I_{3.4\mu{\rm m}} /I_{3.3\mu{\rm m}}$ band ratio (green) according to the distance to the star. The transparent grey box on the left represents the distance up to which the cavity is extended in the POLARIS simulations ($\sim 0.12$" or 13~au).}
    \label{fig_rohsa_profil}
\end{figure}

Figure~\ref{fig_rohsa_maps} shows the maps in the 3.3~$\mu$m aromatic and 3.4~$\mu$m aliphatic emission bands and Fig.~\ref{fig_rohsa_profil} shows their average spatial emission profiles according to the distance $d$ of each pixel from the star.
These two emission bands, due to transient emission after UV photon absorption, may trace the surface of the disk where most of the UV energy is dissipated and converted into near- to mid-IR emission by dust. The carriers being stochastically heated, the band intensities are proportional to the strength of the far-UV radiation field given by $G_0$ expressed in units of the average interstellar radiation field, $1.6 \times 10^{-3}$ erg s$^{-1}$ cm$^{-2}$ \citep{Habing1968}.
The band emission is thus expected to be much more spatially extended than the IR thermal emission of big grains at thermal equilibrium. 
The maps and the spatial profiles show that the emission in the 3.3 and 3.4~$\mu$m bands is in fact spatially extended up to $\sim 1$" (or $\sim 110$~au) and for both bands. We observe an emission peak between 0.2" and 0.4", or 22 and 44~au, located just after the inner edge of the outer disk at about 10-15~au. The decrease in emission for $d > 0.4$" roughly follows a $1/d^2$ law (dashed lines in Fig.~\ref{fig_rohsa_profil}), which corresponds to the dilution law of the far-UV radiation field strength, $G_0$. 

Correlated intensities of aromatic and aliphatic features argue in favour of a common nature of the carriers: stochastically heated nano-particles \citep[see also ][]{Bouteraon2019}.
One can see that, beyond 0.4", the intensity of the 3.3~$\mu$m aromatic band decreases slightly more rapidly than that of the 3.4~$\mu$m aliphatic band. This could reflect size/composition changes in the nano-particles.
Closer to the star, between 0.2" and 0.4", the emission is flat and does not vary anymore with $G_0$. This can be due to disk structure effects, e.g. cavity effects, shadows at the inner edge of the outer disk, or changes in the properties of the nano-particles, e.g. abundance and size.
At large scale, the structure seen in the nano-dust emission appears to match the elliptical structure detected in the stellar light scattered by (sub-)micronic grains \citep[e.g.][]{Augereau2001}.
This is consistent with the expectation that the nano-dust emission traces the disk surface. 

The maps in the aromatic and aliphatic bands show also slightly different morphologies (Fig.~\ref{fig_rohsa_maps}). Figure~\ref{fig_rohsa_profil} shows the average spatial profiles of the aliphatic-to-aromatic band ratio, $I_{3.4\mu {\rm m}}/I_{3.3\mu {\rm m}}$, according to the distance to the star. 
The $I_{3.4\mu {\rm m}}/I_{3.3\mu {\rm m}}$ ratio exhibits a clear increase by a factor of up to 2 when moving away from the star (from $\sim 0.2$" to 1"), which probably reflects the UV processing of the carbonaceous materials. This tendency was visible in Fig.~5 of \citet{Bouteraon2019} but it was not as clear, especially in the disk inner part. This can be explained by the fact that  {\tt ROHSA} is more robust to recover the signal. This will be discussed in more detail in Sect.~\ref{sec_discussion}.

\subsection{VISIR observations}

\subsubsection{Observations}

VISIR observations were performed using the ESO mid-IR instrument VISIR installed on the VLT (Paranal, Chile), equipped with a DRS (former Boeing) 256 $\times$ 256 pixels BIB detector. The object was observed in the imaging and spectroscopic modes. It was observed in the aromatic bands at 8.6 and 11.3~$\mu$m (pah1 and pah2 filters) and in the adjacent continuum at 10.4~$\mu$m. 
Under good seeing conditions it provides diffraction-limited imaging and spectroscopy in the N and Q-band, which is 0.3" at 10~$\mu$m. The spectrometer offers a range in spectral resolution of 150 to 30\,000 and it has a pixel-scale of 127~mas/pix. In order to get rid of the high atmospheric background, the instrument employs standard chopping and nodding techniques.

The observations were obtained in 2005 as part of the VISIR GTO program on circumstellar disks.
The medium resolution spectroscopic mode of VISIR was used. VISIR offers a choice in slit-width, slit-rotation and chopping throw. The orientation of the slit was aligned along the major axis of the disk. For all the observations chopping and nodding was performed parallel to the slit. The standard star used for photometric calibration was HD91056 \citep{Cohen1999}.

\subsubsection{Results}

\begin{figure}[htbp]
    \resizebox{\hsize}{!}
    {\includegraphics[width=\textwidth,height=\textheight,keepaspectratio]{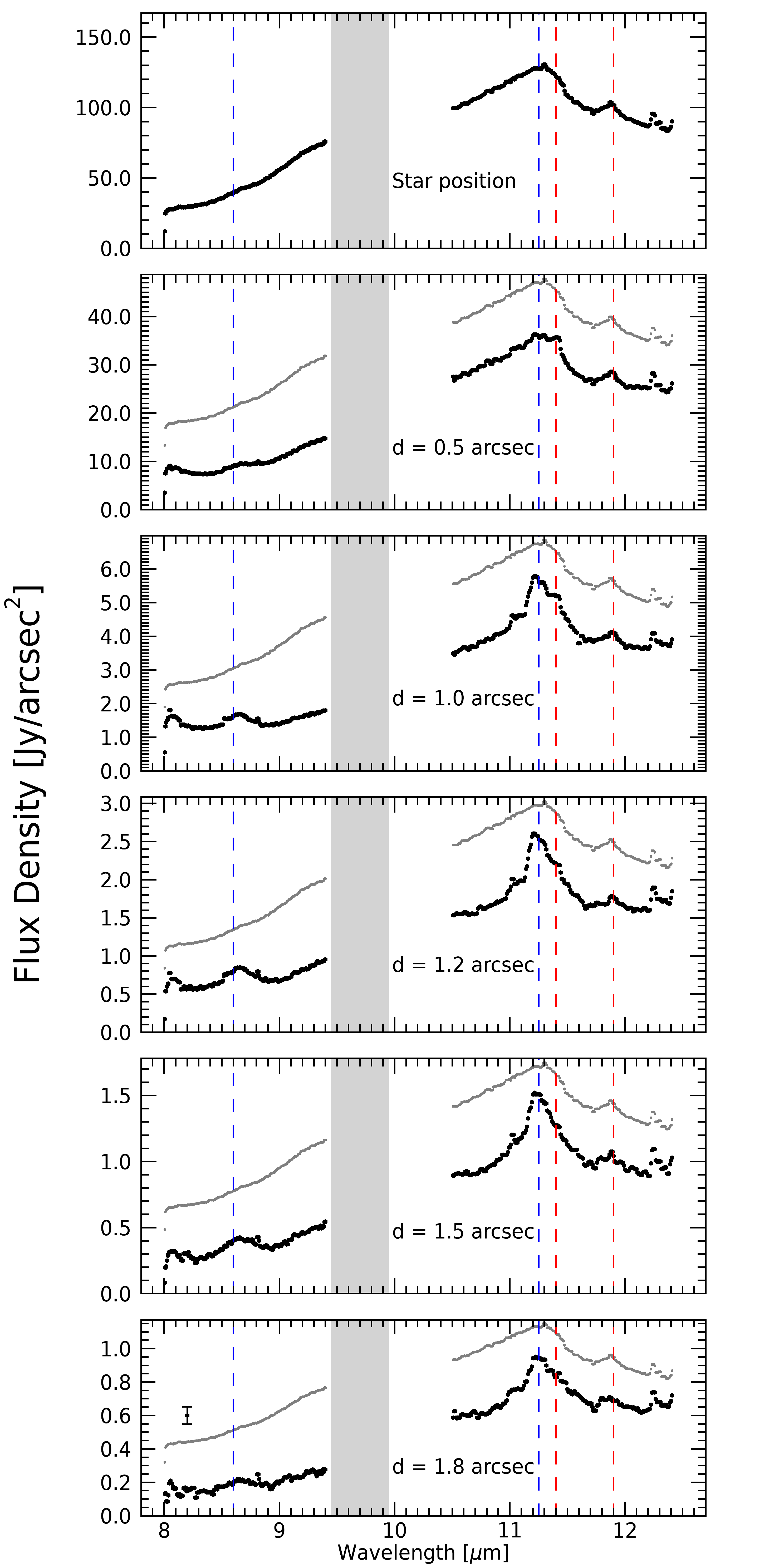}}
    \caption{
    VISIR observed spectra in HD~100546 extracted at different angular distances (0, 0.5, 1.0, 1.2, 1.5 and 1.8''). The grey region delimits the telluric ozone absorption feature. The corresponding grating set-up has not been recorded at that position due to the usual bad quality of the data. A rescaled/offset star position spectrum is repeated (light grey) on all panels to serve as a reference. Overplotted in dashed lines are the positions of the aromatic (blue) and crystalline forsterite (red) main features in VISIR observation range. The typical $\pm 3\sigma$ error derived from background noise is shown in the lower panel.
    }
    \label{fig_visir_obs}
\end{figure}

Figure~\ref{fig_visir_obs} shows VISIR mid-IR spectra measured at different distances from the central star (up to 1.8"). A rescaled/offset star position spectrum is shown (light grey) in all panels to serve as a reference. These spectra are centred on the 8.6 and 11.3~$\mu$m aromatic features, which correspond to aromatic C-H bending in-plane and out-of-plane vibrational modes, respectively (see Tab.~\ref{tab_versat}). In Figure~\ref{fig_visir_obs} overplotted in dashed lines are the positions of the aromatic and crystalline silicate main features.

The 8.6~$\mu$m feature is not observed in the spectra at the star's location but its intensity relatively to the adjacent continuum increases significantly from 0.5" to 1.5". 
At 1.8", its intensity relatively to the  continuum appears to be lower than at 1.5". In the 11-12~$\mu$m range of the spectra, two features at 11.4 and 11.9~$\mu$m, characteristic of crystalline silicates, are clearly detected at the star location and up to a large distance from the star (at least 1.5").
The emission in the 11.3~$\mu$m band, relative to the emission in the crystalline silicate band, increases significantly between 0.5" and 1.5". At 1.8", weak aromatic and silicate bands appear. We checked that the influence of the contribution from the unresolved peak component on the spectra is not significant at large distances. This contribution is in fact estimated to be at most 20\% at 1.5-1.8".

\begin{figure}[htbp]
    \resizebox{\hsize}{!}
    {\includegraphics[width=\textwidth,height=\textheight,keepaspectratio]{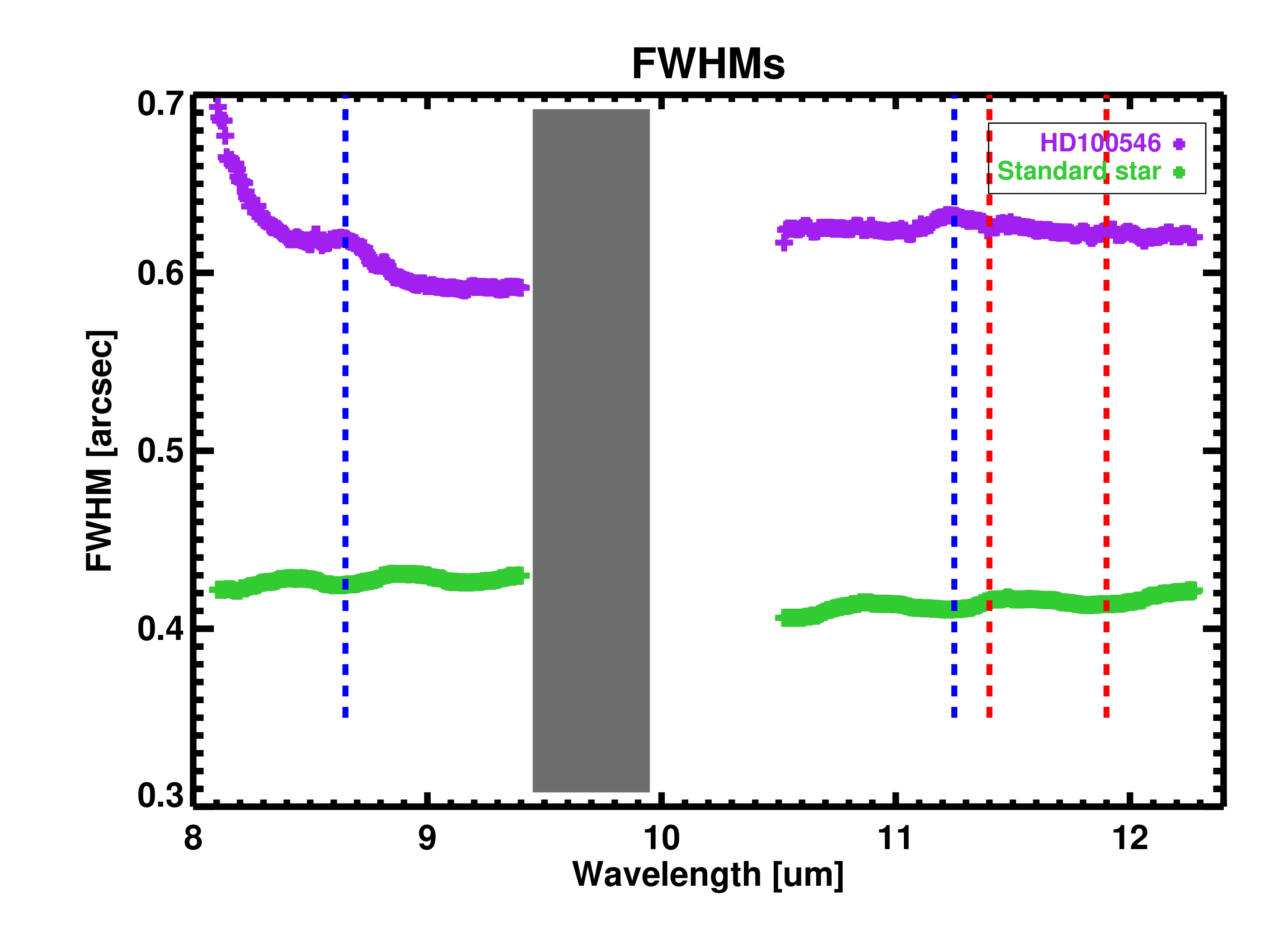}}
    {\includegraphics[width=9cm,height=9cm,keepaspectratio]{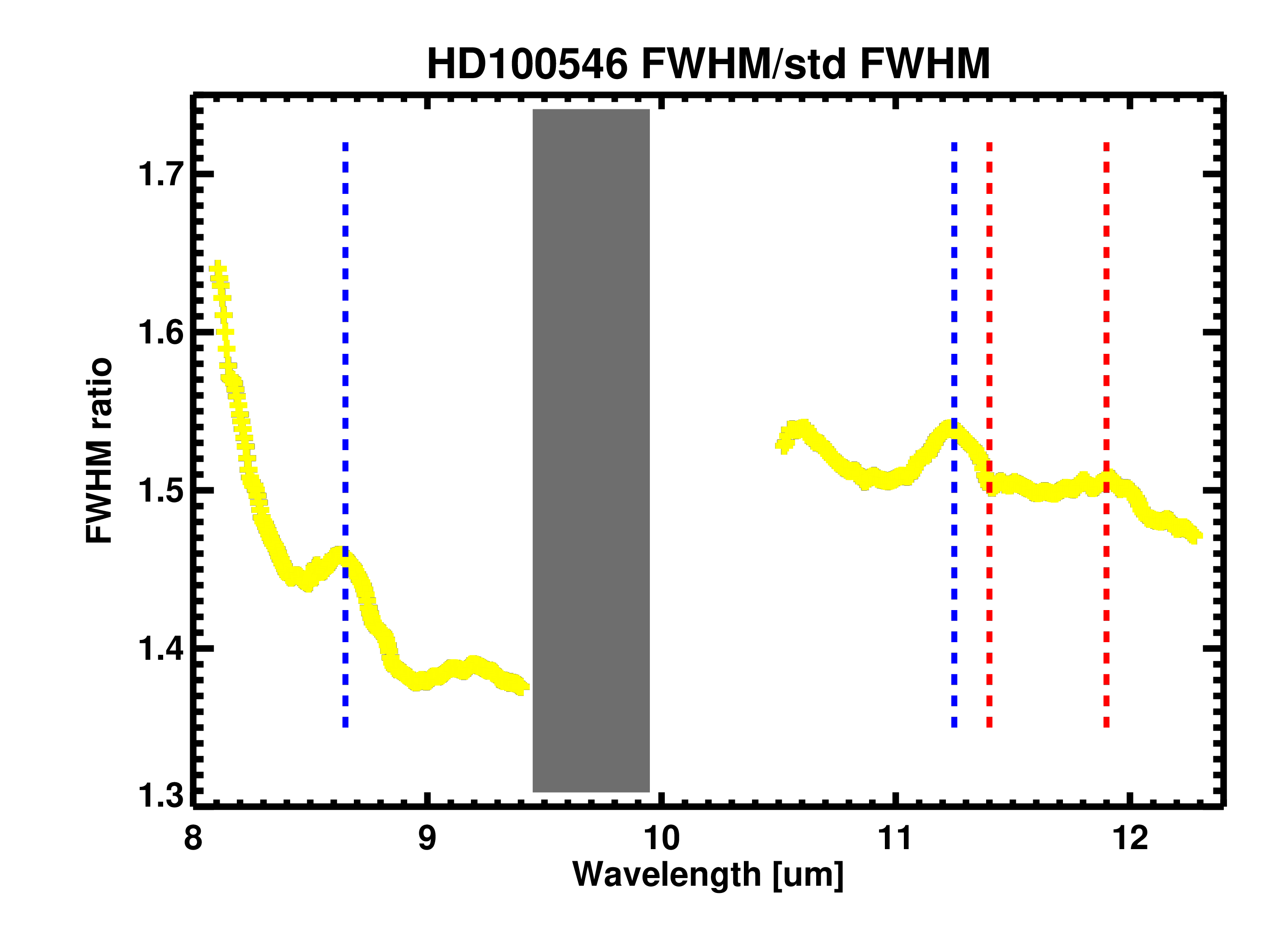}}
    \caption{
Upper figure: measured FWHM in VISIR long-slit spectra. The ozone atmospheric absorption band is delimited by the grey shaded area. 
The positions of aromatic (blue) and silicate crystalline (red) main features are shown as dashed lines. The estimated 3 $\sigma$ error on the FWHM is 2.5~10$^{-3}$".
The purple points are the measures on HD100546. The same reference star (HD91056, green points) as the one used for photometric calibration has been used and is supposed to reflect the angular resolution response of the atmosphere combined to the telescope/instrument.
Bottom figure: Ratio between HD100546 and standard star FWHM shown in the upper figure. The raw ratio obtained by crude division of the FWHMs has been smoothed it with a 3-pixels wide boxcar having the spectral width of the actual spectral sampling  in order to reduce pixel-to-pixel irregularities. The estimated 3 $\sigma$ error on the ratio is about 1.2\% ($\sim$0.02).
}
    \label{fig_visir_FWHM_obs}
\end{figure}

In order to measure the spatial scales of the bands, the VISIR long-slit spectra have been Gaussian-fitted following \citep[]{Boekel2004PhD} procedure to 
obtain the FWHM-sizes as a function of wavelength (see Figure~\ref{fig_visir_FWHM_obs}). At each wavelength, the spatial profile is adjusted using a 1D gaussian model. From the gaussian model, a FWHM is then derived. Figure~\ref{fig_visir_FWHM_obs} shows the measured FWHM on HD100546 and the standard star, as well as, the ratio between HD100546 and standard star FWHM. The standard star is assumed to reflect the angular resolution response of the atmosphere combined to the telescope/instrument. The FWHM of the spatially resolved spectrum of HD100546 is systematically increased in the aromatic bands. The 11.3~$\mu$m has a larger FWHM than that of 8.6~$\mu$m. The 7.7~$\mu$m appears even more extended, although we only see the rise. The FWHM of HD100546 is also larger at the 11.9~$\mu$m crystalline silicate feature than the adjacent continuum. We do not see anything special at 11.4~$\mu$m, but it is blended with the aromatic 11.3~$\mu$m band. The FWHM increase in the crystalline band may suggest the presence of nano-silicates warm far from the star.

\begin{figure}[htbp]
    \resizebox{\hsize}{!}{\includegraphics[width=\textwidth,height=\textheight,keepaspectratio]{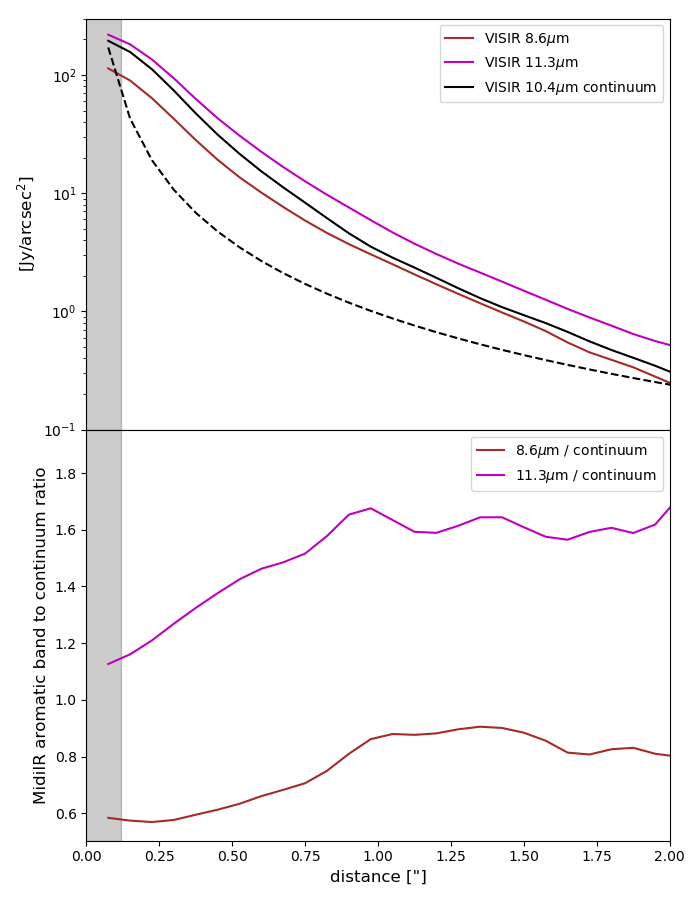}}
    \caption{Top panel: the 8.6~$\mu$m (brown) and 11.3~$\mu$m (magenta) aromatic emission bands, and the 10.4~$\mu$m (black) continuum emission are plotted according to the distance to the star. The $1/d^2$ function is also plotted (black dotted). Bottom panel: the 8.6~$\mu$m/10.4~$\mu$m and 11.3~$\mu$m/10.4~$\mu$m band ratios are plotted according the distance to the star. The transparent grey box on the left represents the distance up to which the cavity is extended in POLARIS simulations, $\sim 0.12$" (13~au). At large distances, the variations in the band-to-continuum ratios are not significant. The error bars on the band-to-continuum ratio are about 5\%. }
    \label{fig_visir_profil}
\end{figure}

Fig.~\ref{fig_visir_profil} shows the spatial emission profiles in the VISIR filters centred on the aromatic bands at 8.6~$\mu$m and 11.3~$\mu$m and on the 10.4~$\mu$m continuum, as well as the band-to-continuum ratios, depending on the distance from the star. We note that the emission profiles in the VISIR filters are not continuum-subtracted which is different from Fig. \ref{fig_rohsa_profil} showing the fluxes in the bands.
Fig.~\ref{fig_visir_profil} shows that both the emission centred on the bands and on the continuum are spatially extended.
The two aromatic band-to-continuum ratios increase with the distance from the star up to 1" ($\sim 100$~au) and are constant at a larger distances. 
The low ratio values closer to the star are mostly due to the increase in the thermal emission of the larger grains at thermal equilibrium, the emission of which peaks at shorter wavelength when the UV flux increases.
Up to a distance of 0.2 to 0.3" (or $\sim 20-30$~au) from the star where $10^6 \lesssim G_0 \lesssim 10^7$, sub-micronic grains can reach an equilibrium temperature of $\sim 300$~K and thus strongly emit in the 10~$\mu$m range. 
On the other hand, the aromatic band profiles decrease slightly faster than a $1/d^2$ law (dashed line in Fig.~\ref{fig_visir_profil}), showing that emission in the two bands does not simply scale linearly with $G_0$, a behaviour expected if emission is dominated by a single population of stochastically heated grains. This means that the emission could be due to a mix of the contributions of different nano-grain populations, e.g. extremely small to large nano-particles. As a consequence, the large nano-particles reach thermal equilibrium for the high $G_0 > 10^4$ values at the disk surface and therefore modify the scaling of their mid-IR emission spectrum (see Sect.~\ref{sec_polaris} for details).
However, near the star, the emission profiles are dominated only by the thermal hot grain emission.

These observations also indicate that the mid-IR aromatic features are spatially more extended than the 3.3~$\mu$m aromatic and 3.4~$\mu$m aliphatic ones (Sect.~\ref{subsub_rohsa-res}). Among the aromatic bands, the less energetic ones (at 8.6 and 11.3~$\mu$m) are in fact expected to come mostly from the outer disk region \cite[e.g.][]{Habart2004}, whereas the most energetic one (at 3.3 $\mu$m) is expected to be the less extended.

In order to probe the properties of the emitting particles, one could compare the intensity band ratios, $I_{11.3\mu{\rm m}}/I_{8.6\mu{\rm m}}$, particularly sensitive to their ionisation state, as well as, the $I_{11.3\mu{\rm m}}/I_{3.3\mu{\rm m}}$, mostly depending on their size \citep[e.g.][]{croiset2016}.
Nevertheless, from 0.5 to 1.5", where the  aromatic bands are detected, doing such an analysis with our NaCo and VISIR data would be difficult. Indeed, the 8.6~$\mu$m band is partly hidden by the strong continuum whereas the 11.3~$\mu$m band is blended with the crystalline silicate feature at 11.4~$\mu$m (see Fig.~\ref{fig_visir_obs}). Such an analysis is thus beyond the scope of this study but will be of prime interest when high signal-to-noise observations with the James Webb Space Telescope (JWST) will become available. In particular, the emission at 8.6 and 3.3~$\mu$m will be detectable in the outer disk regions.

\section{Disk modelling with THEMIS and POLARIS \label{sec_polaris}}

\subsection{THEMIS dust model \label{subsec_themis}}

The Heterogeneous dust Evolution Model for Interstellar Solids (THEMIS\footnote{\url{https://www.ias.u-psud.fr/themis/}}) 
is an evolutionary core-mantle dust model designed to allow variations in the dust structure, composition, and size according to the local density and radiation field \citep{Jones2013, Jones2014, Bocchio2014, Koehler2015, Jones2017}. THEMIS has been used before to analyse far-IR to submm observations of the diffuse and dense ISM \citep[e.g.][]{Ysard2015, Ysard2016}, near-IR PPD observations \citep{Bouteraon2019}, and near-IR to submm observations of nearby galaxies \citep{Chastenet2017,viaene2019}.

In the following, unless otherwise stated, we use the dust composition and size distribution of the THEMIS diffuse-ISM dust defined by \citet{Jones2013, Koehler2014} (Tab.~\ref{tab_dust_parameter}). It consists in three dust populations: a first one made of small hydrogenated amorphous carbon nano-particles ($< 20~$nm, a-C); a second one of larger carbonaceous grains with an hydrogen-rich core surrounded by a 20~nm thick aromatic-rich mantle (a-C:H/a-C); and a third one of large amorphous silicate grains of which 10\% of the volume is occupied by metallic nano-inclusions of Fe and FeS (a-Sil/a-C). In addition, these silicate cores are surrounded by a 5~nm thick a-C mantle. We thus consider diffuse-ISM like grains with radius $a \lesssim 1~\mu$m. Since, at the disk surface, these small grains absorb  most of the stellar UV/visible light and dominate the emission in the near- to mid-IR, this is expected to provide a consistent modelling of the NaCo and VISIR observations of HD~100546.

In THEMIS, the population of carbon grains a-C with sizes from 0.4 to 20~nm has continuous size-dependent optical properties as well as a continuous size distribution contrary to most dust models \citep[e.g.][]{Desert1990, draine_infrared_2001, Compiegne2011}.

The grain structure and optical properties are thus calculated by considering their composition and size, and the resulting spectra are in direct relationship with these properties.
In order to ease the comparison with previous models and studies, we distinguish three sub-populations of a-C grains:
\begin{itemize}
    \item a-C1: a-C nano-particles with $a \leq 0.7$~nm which would be equivalent to the smallest PAH grains in \citet{Desert1990, draine_infrared_2001, Compiegne2011} and represent more or less $\sim 10$\% of the mass for all models;
    \item a-C2: a-C with $0.7 < a \leq 1.5$~nm which would be equivalent to the largest PAHs in \citet{Desert1990, draine_infrared_2001, Compiegne2011} ;
    \item Very Small Grains (VSG): larger a-C particles with $a > 1.5$~nm which would be equivalent to the VSG as defined in \citet{Desert1990}, the smallest carbon grains with graphite optical properties in \citet{draine_infrared_2001}, and the Small Amorphous Carbon grains (SamC) in \citet{Compiegne2011}.
\end{itemize}

The dust mass fraction of each THEMIS dust population and a-C sub-population, i.e. the population dust mass $M_X$ to the total dust mass $M_{total}$ ratio, $q_X = M_X/M_{total}$ are:
\begin{itemize}
    \item $q_{{\rm a-C1}} \sim 10\%$;
    \item $q_{{\rm a-C2}} \sim 7\%$;
    \item $q_{{\rm VSG}} \sim 6\%$;
    \item $q_{{\rm a-C:H/a-C}} \sim 8\%$;
    \item $q_{{\rm a-Sil/a-C}} \sim 69\%$.
\end{itemize}

Finally, we have to mention that for simplicity, we consider only amorphous silicate grains in our model, while several signatures from crystalline silicates have clearly been observed towards HD~100546 as shown in Fig.~\ref{fig_visir_obs} and in previous studies \citep[e.g.][]{Malfait1998,Bouwman2003,sturm2010.}
However, this assumption will not affect our results, since we focus on the spatial distribution of the nano-carbon grain emission features and do not do a detailed analysis of the band profiles.

\begin{table*}
	\caption{Standard THEMIS model parameters as described in \citet{Jones2017}. 
	}
	\label{tab_dust_parameter}
	\centering
	\begin{tabular}{ccccc}
	    \hline \hline 
Dust population & Size distribution type & Characteristic size (nm) & Density (g/cm$^{3}$) & $M_X$/ $M_{total}$ (\%) \\
\hline
  a-C1      & power law   & 0.4-0.7     & 1.6  & $\sim$10 \\
  a-C2      & power law   & 0.7-1.5     & 1.6  & $\sim$7  \\
  VSG       & power law   & 1.5-20      & 1.6  & $\sim$6  \\
  a-C:H/a-C & log-normal  & $\sim 160$  & 1.57 & $\sim$8  \\
  a-Sil/a-C & log-normal  & $\sim 140$  & 2.19 & $\sim$69 \\
\hline
\end{tabular}
\label{themis_tab_vs}
\end{table*}

\subsection{POLARIS radiative transfer code \label{subsec_disk}}

In order to analyse the NaCo and VISIR data with THEMIS, we apply the three-dimensional continuum radiative transfer code POLARIS \citep[][]{Reissl2016, brauer_magnetic_2017-1}. It solves the radiative transfer problem self-consistently with the Monte Carlo method. Main applications of POLARIS are the analysis of magnetic fields by simulating polarised emission of elongated aligned dust grains and the Zeeman splitting of spectral lines. However, recent developments added circumstellar disk modelling as a new main application to POLARIS \citep[e.g.,][]{Brauer2019}. For instance, this includes the consideration of complex structures and varying dust grain properties in circumstellar disks as well as the stochastic heating of nanometre-sized dust grains. We use POLARIS to calculate the spatial dust temperature distribution, emission maps as well as spectral energy distributions based on the thermal emission of dust grains, the stochastic heating of nanometre-sized dust grains and the stellar emission scattered by dust grains.
The code was published on the POLARIS homepage\footnote{\url{http://www1.astrophysik.uni-kiel.de/~polaris/}}. For reference, the simulations in this study were performed with the version 4.02. of POLARIS.

Our theoretical model of the circumstellar disk consists of a pre-main sequence star in the centre, which is surrounded by an azimuthally symmetric density distribution. The star is characterised by its effective temperature $T_\text{star} = 10\,500$~K and luminosity $L_\text{star}=\SI{32}{L_\odot}$. For the disk density, we consider a distribution with a radial decrease based on the work of \citet{hayashi_structure_1981} for the minimum mass solar nebula. Combined with a vertical distribution due to hydrostatic equilibrium similar to the work of \citet{Shakura1973}, we obtain the following equation:
\begin{equation}
	\rho_\text{disk}=\rho_0 \left( \frac{R_\text{ref}}{r} \right)^{a} \exp\left(-\frac{1}{2}\left[\frac{z}{H(r)}\right]^2 \right).
	\label{eqn:disk}
\end{equation}
Here, $r$ is the radial distance from the central star in the disk midplane, $z$ is the distance from the midplane of the disk, $R_\text{ref}$ is a reference radius, and $H(r)$ is the scale height. The density $\rho_0$ is derived from the disk mass. The scale height $H(r)$ is a function of $r$ as follows:
\begin{equation}
	H(r)=h_0 \left(\frac{r}{R_\text{ref}}\right)^{b}.
	\label{eqn:disk2}
\end{equation}
The parameters $a$ and $b$ set the radial density profile and the disk flaring, respectively. $h_0$	is the scale height at the characteristic radius $R_{ref}$=100~sau.  
\citet{Benisty2010} found for their best disk model around HD~100546 a scale height of 10~au at a distance of 100~au, which is equivalent to a flaring angle of 7$^{\circ}$. In their model, they assumed a standard flaring index ($b = 1.125$). \citet{Avenhaus2014} suggests that the disk could be more strongly flared in the outer regions.
In our model, we consider a scale height of 10~au at 100~au, similar to \citet{Benisty2010}. For the disk flaring $b$ we assume $b=9/7 \sim 1.28$ as expected from hydrostatic radiative equilibrium models of passive flared disks \citep{chiang1997}.
The extent of the disk is constrained by the inner radius $R_\text{in}$ and outer radius $R_\text{out}$. 
Following previous studies on HD100546 discussed in Sect.\ref{sec_HD100546}, we consider a disk model with 
an inner radius of 0.2 au, an outer radius of 350 au, and a cavity between 5 and 13 au \citep[e.g.][]{Bouwman2003,Brittain2009,Benisty2010, Tatulli2011, Mulders2013,Panic2014, Garufi2016, Follette2017}.
Moreover, the inclination of the disk is taken equal to 42$^{\circ}$ \citep{Ardila2007}.
An overview of the assumed parameters of our disk model can be found in Tab.~\ref{tab_model_parameter}. 
In the model, we assume a dust-to-gas mass ratio of $M_{dust}/M_{H}\sim 7 \times 10^{-4}$, e.g. 10 times lower than the standard THEMIS model for diffuse ISM.

The dust-to-gas mass ratio is low compared to common literature values (e.g., $\sim\SI{0.01}{}$), but we consider in our simulations only the small grains with $a \lesssim 5~\mu$m, while due to grain growth in disks a significant amount of the mass is present in larger grains.
However, since the disk is optically thick in the NaCo and VISIR wavelength range, 
the predicted dust emission profiles are not sensitive to the non-included larger grains, but rather to the relative mass fraction between the different nanometer- and micron-sized grain populations.

\begin{table}
	\caption{Parameters of the circumstellar disk model. Only small grains with $a \lesssim 5~\mu$m are considered. 
	References : 
    (1) \citet{acke_iso_2004};
	(2) \citet{Gaia2018};
	(3) \citet{Benisty2010}, (4) \citet{Tatulli2011}, (5) \citet{Mulders2013}, (6) \citet{Panic2014};  
	(7) \citet{Augereau2001,Benisty2010};
	(8) \citet{Benisty2010};
	(9) \citep{Ardila2007}.
	}
	\label{tab_model_parameter}
	\centering
	\begin{tabular}{llll}
	    \hline \hline 
		\multicolumn{2}{l}{Parameter} & Value  & Ref.                                                        \\
		\hline 
		\multicolumn{3}{c}{\textit{Central star}}                                                                   \\
		\hline 
		Effective temperature                & $T_\text{star}$              & $\SI{10500}{K}$     & (1)                  \\
		Stellar luminosity                   & $L_\text{star}$              & $\SI{32}{L_\odot}$  & (1)                  \\
		\hline 
		\multicolumn{3}{c}{\textit{Disk model}}                                                                     \\
		\hline 
		Distance          & $d$                & 110 pc & (2)\\
		Gas mass                  & $M_{gas}$          & $10^{-2}M_\odot$ \\
		Dust-to-gas mass ratio    & $M_{dust}/M_{gas}$ & 7.43e-4 \\
		Inner radius              & $R_{in}$           & 0.2 au & (3,4,5,6)\\
		Outer radius              & $R_{out}$          & 350 au & (7) \\
		Cavity                    &                    & 5-13 au & (8)\\
		Scale height              & $h_0$              & 10 au \\
	    Characteristic radius     & $R_{ref}$          & 100 au \\
		Radial density exp.       & $a$                & 2.357 \\
		Disk flaring exp.         & $b$                & 1.286 \\
		Cells in $r$-direction    & $n_r$              & 100 \\
		Step width factor in $r$  & sf                 & 1.03 \\
		Cells in $z$-direction    & $n_\theta$         & 142 \\
		Cells in $\phi$-direction & $n_\phi$           & 1 \\
		Disk inclination          &                    & 42$^{\circ}$ & (9)\\

		\hline
	\end{tabular}
\end{table}

\subsection{Model results}

\subsubsection{Physical conditions in the upper surface layers}

\begin{figure*}[htbp]
    \centering
    \includegraphics[width=17cm]{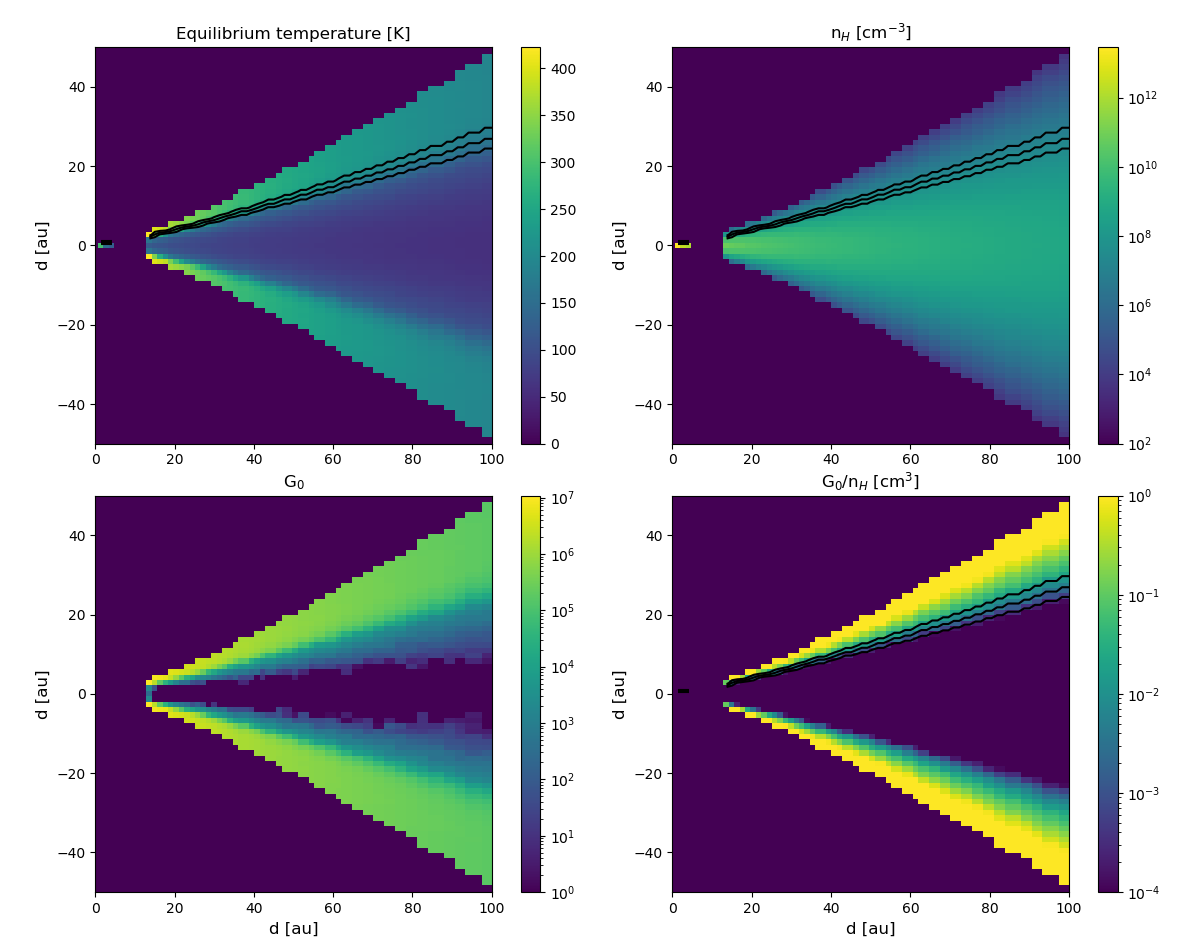}
    \caption{Vertical cuts through the model disk from our POLARIS simulation. The cuts show the temperature (K) distribution of large a-C(:H)/a-C at thermal equilibrium (\textit{top left}), the hydrogen density $n_H$ (cm$^-3$) distribution (\textit{top right}), the radiation field intensity $G_0$ (\textit{bottom left}) and the $G_0/n_H$ (cm$^3$) ratio (\textit{bottom right}) of the two previous maps. The black lines correspond to 25, 50, and 75\% of the total emission at 3.3~$\mu$m seen from the top. The white area in the two lower panels is due to the fact that $G_0$ is not defined in the most median parts of the disk because UV radiation does not penetrate.}
    \label{fig_1_maps}
\end{figure*}

Figure~\ref{fig_1_maps} shows maps of the temperature distribution of large a-C:H/a-C at thermal equilibrium, and of the hydrogen density n$_H$, radiation field intensity $G_0$, and $G_0/n_H$ ratio distributions from our disk model. 
The black lines correspond to 25, 50, and 75\% of the total emission at 3.3~$\mu$m seen from the top. The emission at 3.3~$\mu$m is completely dominated by the upper disk layers due to the high optical depth of the model disk. 

\begin{figure}[htbp]
  \includegraphics[width=9cm]{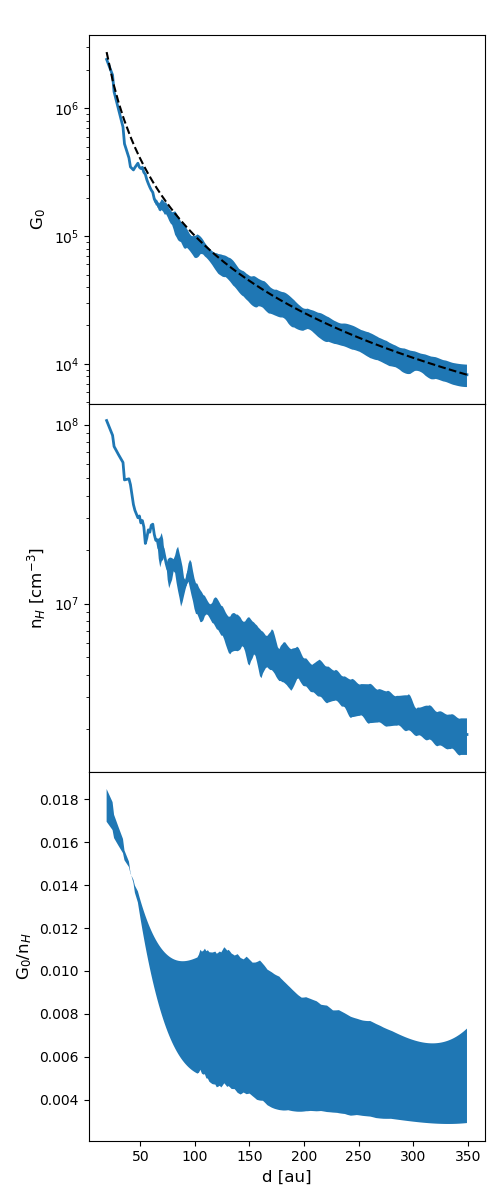}
    \caption{$G_0$, $n_H$, $G_0/n_H$ according to distances to the star for the model. $G_0$ is the average of $G_0$ values between 25 and 75\% of the total emission at 3.3~$\mu$m seen by the observer (Fig.~\ref{fig_1_maps}, bottom left panel). Idem for n$_H$ (top right panel) and  G$_0$/n$_H$ (bottom left). The width of the line is for the standard dispersion of the values. $1/d^2$ (black dotted line) is also plotted in the top panel.}
    \label{fig_polaris_g0}
\end{figure}

Figure~\ref{fig_polaris_g0} presents $G_0$, $n_H$, and $G_0/n_H$ according to the distance to the star for the disk layers where 25 to 75\% of the total emission at 3.3~$\mu$m are seen by the observer. 
The radiation field $G_0$, which varies as 1/d$^2$, decreases by a factor of about 300 from 20~au to 350~au, with a minimum value slightly lower than 10$^4$ at 350~au. 
The gas density $n_H$ decreases by a factor of 100 from 20~au to 350~au, with a minimum value of $\sim 2 \times 10^6~$~cm$^{-3}$ at 350~au.
Then, as both $G_0$ and $n_H$ decrease with the distance, the $G_0 / n_H$ ratio does not vary much to first order in the disk layer from which most of the a-C band emission comes. The $G_0/n_H$ ratio decreases by a factor 3 from $\sim 0.018$ at 20~au to $\sim 0.006$ at 100~au and remains roughly constant up to 350~au. 

\subsubsection{Predicted spectra between 3 and 20~$\mu$m}

\begin{figure}[htbp]
\resizebox{\hsize}{!}{\includegraphics[width=9cm,height=\textheight,keepaspectratio]{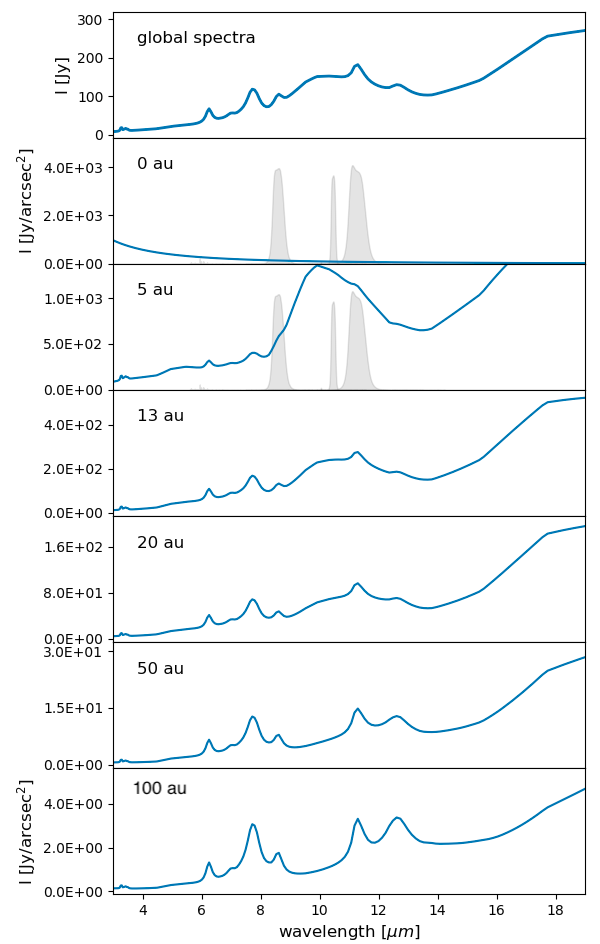}}
    \caption{IR emission POLARIS spectra between 3 and 20~$\mu$m. Global (Jy) (top panel) and individual emission spectra (Jy/arcsec$^2$) are plotted from 1~au (second panel) to 100~au (bottom) from the star.
    VISIR filters are also plotted (grey). }
    \label{fig_1_spec_2}
\end{figure}

\begin{figure*}[htbp]
    \centering
    \includegraphics[width=17cm]{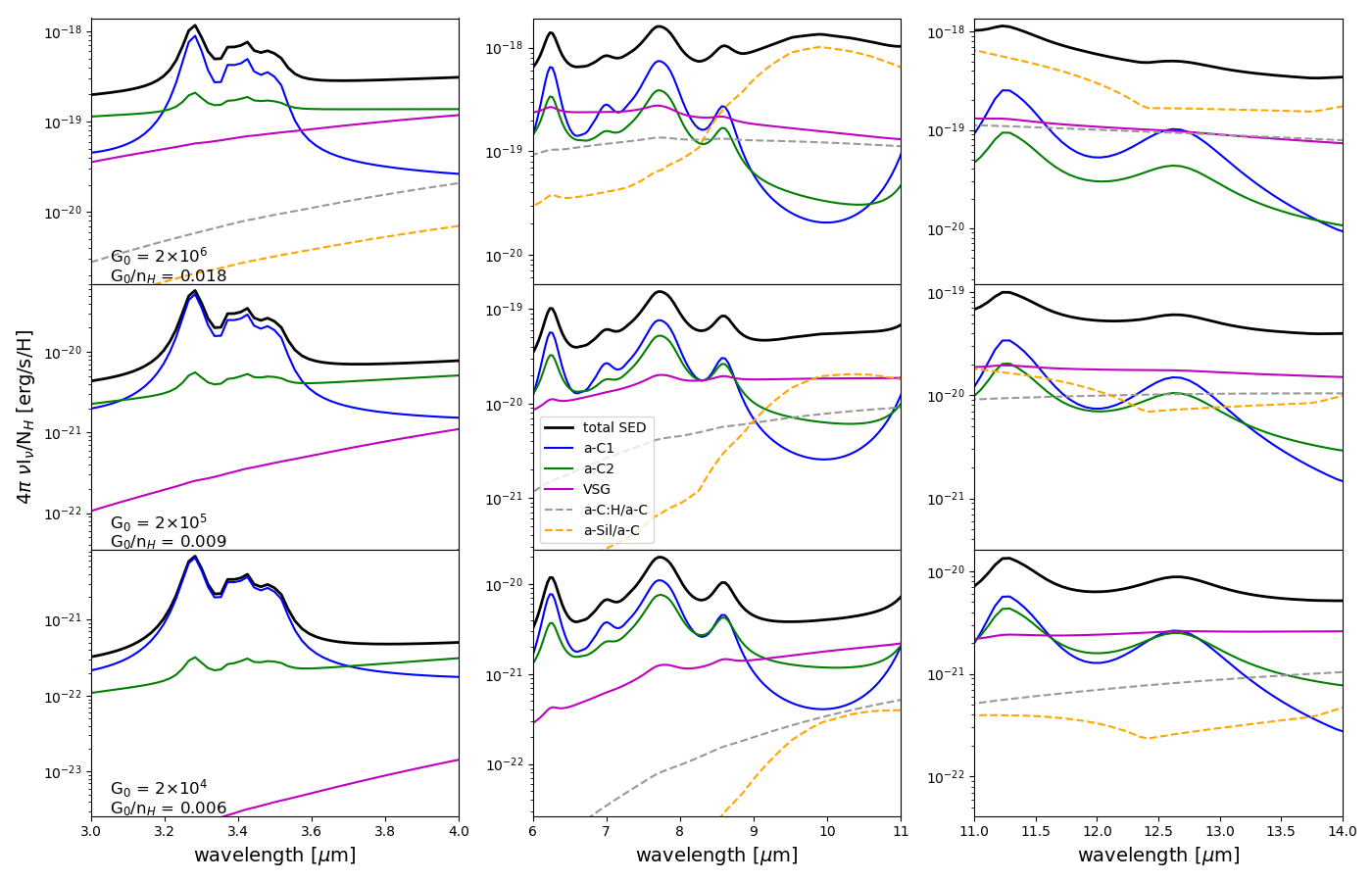}
    \caption{Grain population emission in the IR: total emission (black), a-C1 (blue), a-C2 (green), VSG (magenta), a-C:H/a-C (grey dot), and a-Sil/a-C (orange dot). The left panels are for the NaCo range between 3 and 4~$\mu$m, the middle and right panels correspond to the range covered by VISIR filters: 8.6, 10.4, and 11.3 $\mu$m.
    From top to bottom: decrease in the $G_0$ values which are $2 \times 10^6$, $2 \times 10^5$, and $2 \times 10^4$. The corresponding distances to the star are  20, 75, and 200~au, respectively. The distance and $G_0 / n_H$  are derived from Fig.~\ref{fig_polaris_g0}.}
    \label{fig_demix_sed}
\end{figure*}

Here, we present the spectra between 3 and 20~$\mu$m calculated for the disk model using POLARIS and THEMIS as a function of the distance from the star, from 1 to 100~au (see Fig. \ref{fig_1_spec_2}).
Predicted spectra are different in the inner and outer disk parts, as expected.
In order to ease the analysis of the predicted spectra, based on THEMIS and DustEM calculations, Fig.~\ref{fig_demix_sed} shows the emission of each dust sub-population between 3 and 14~$\mu$m for several distances to the star. 

Fig.~\ref{fig_demix_sed} shows that between 3 and 4~$\mu$m, the emission in the aromatic and aliphatic bands is highly dominated by the smallest a-C nano-particles (a-C1 \& a-C2) and the contribution of the bigger VSG grains is negligible ($a > 1.5$~nm). When the radiation field intensity $G_0$ decreases, only the emission due to the smallest a-C nano-particles, a-C1, remains. At large distance from the star, both the continuum and band emissions are due to these extremely small nano-particles ($a < 0.7$~nm). 
For high $G_0$ values, the contribution of large silicates and VSG grains is significant in the VISIR 8.6~$\mu$m bandpass. When the radiation field intensity $G_0$ decreases, the band emission from $\sim 6$ to 9~$\mu$m is due both to a-C1 and a-C2 amorphous carbon nano-particles. 
The shape of the 11.3~$\mu$m band is strongly affected by the emission of big silicates for high $G_0$ values. When the radiation field intensity $G_0$ decreases, the band emission is then dominated by a-C nano-particles emission, whatever their sizes (a-C1, a-C2, VSG) are. 

All aromatic bands are thus due to different combinations of grain sub-populations with a high dependence on the intensity of the radiation field:
\begin{itemize}
    \item 3.3~$\mu$m: a-C1 at both at low and high $G_0$;
    \item 8.6~$\mu$m: a-C1 and a-C2 at low $G_0$, a-C1, a-C2, VSG and a-Sil/a-C at high $G_0$;
    \item 11.3~$\mu$m: a-C1 and a-C2 at low $G_0$, a-C1 and a-Sil/a-C at high $G_0$.
\end{itemize}

Looking at the 10.4~$\mu$m continuum, except for the smallest $G_0$, the emission is a mix of the contribution of all grain populations. 
For $G_0 = 2 \times 10^6$ (equivalent to a distance of 20~au), large silicates contribute to $\sim 75$\%, while both large carbonaceous a-C:H/a-C and VSG grains to $\sim 10$\%. For $G_0=2 \times 10^5$ (75~au): both large silicates and VSG grains contribute to $> 30$\%, a-C:H/a-C grains to $\sim 15$\%, a-C2 to $\sim 10$\%, and a-C1 to $\sim 5$\%.
For $G_0 = 2 \times 10^4$ (200~au): VSG contribute to $\sim 45$\%, a-C2 to $\sim 30$\%,  a-C1,  a-C:H/a-C and silicates to $\sim 10$\% or less. However, it should be noted that the SEDs presented in Fig.~\ref{fig_demix_sed} do not take into account the scattered light. In our POLARIS simulations, this scattered light can make up to about half of the continuum in the near-IR but is negligible at wavelengths larger than 10~$\mu$m. This is true for the characteristic silicate grain size of 140~nm presented in our model (see Tab.~\ref{tab_dust_parameter}) and remains valid for sizes up to $\sim 1$~mm.

\section{Comparison between model predictions and observations} \label{sec_model_obs}

In this section, we compare the model predictions to the observations. We first discuss the predicted and the
observed integrated fluxes in the nano-dust bands over the disk, and then the comparison between their predicted and observed spatial emission profiles by NaCo and VISIR as a function of the distance to the central star. 

\subsection{Predicted and observed integrated fluxes in the bands over the disk}
\label{sec_model_obs_1}

\begin{figure}[htbp]
\resizebox{\hsize}{!}{\includegraphics[width=\textwidth,height=\textheight,keepaspectratio]{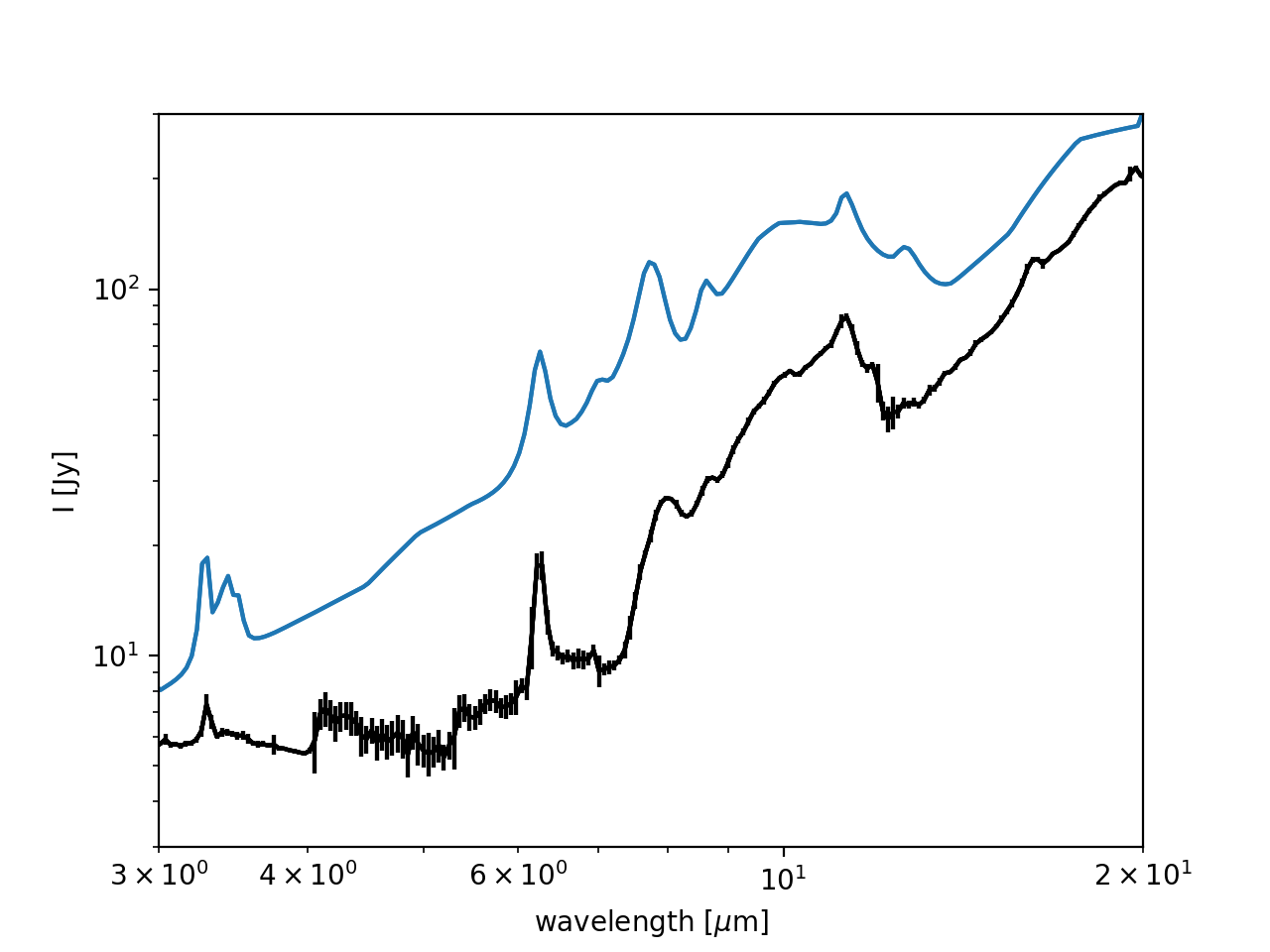}}
\caption{Global spectra (Jy) between 3 and 20~$\mu$m as observed by ISO (black) and as predicted by the model (blue).
}
    \label{fig_SED_modobs}
\end{figure}

Figure \ref{fig_SED_modobs} presents the global spectra over the disk between 3 and 20~$\mu$m as observed and as predicted by the model.
The predicted spectrum reproduces the overall shape of the observed spectrum with the main bands. Nevertheless, the predicted flux is overall higher than that observed by a factor of 2-4.
Table~\ref{tab_flux_bands} presents the integrated fluxes of the aromatic and aliphatic features at 3.3 and 3.4~$\mu$m, as well as, the aromatic features at 6.2 and 8.6~$\mu$m, as observed and as predicted by the model over the disk. For the observation of the aromatic bands, we take the fluxes as derived from the ISO spectrum \citep{Kerckhoven2002phd}. 
We have not included the 11.3~$\mu$m band since it is blended with strong silicate emission. For the observation of the aliphatic band, we take the fluxes as derived from the NaCo spectrum over the disk.
The model overestimates the integrated fluxes in the aromatic band at 3.3~$\mu$m by a factor of 4. This difference is mainly due to the model's overestimation in the internal regions of the disk ($r<40$~au) as suggested by the comparison of the observed and predicted profiles (see next section). 
The band at 3.3~$\mu$m reaches 50\% of its flux at a radius of $r < 20-30$~au (0.18-0.27").  
The aliphatic-to-aromatic band ratio $I_{3.4_{\mu{\rm m}}}/I_{3.3_{\mu{\rm m}}}$ observed over the disk is well reproduced by the models within 20\%.
The model overestimates the integrated flux in the band at 6.2 and 8.6~$\mu$m by a factor of 2.5. This difference is also mainly due to the model's overestimation in the internal regions of the disk. The difference with the observations is a little less important than for the 3.3~$\mu$m because these bands at longer wavelength come from regions more external. The band at 8.6~$\mu$m reaches 50\% of its flux at a radius of $r < 60$~au (0.54"). The model overestimation in the external regions of the disk can also contribute but in less important ways.

\begin{table}
\caption{The integrated fluxes of the aromatic and aliphatic features as observed and as predicted by the model over the disk. Integrated fluxes (after continuum subtraction) and uncertainties are in $10^{-14}$~W/m$^2$. 
}
\label{tab_flux_bands}      
\centering        
\begin{tabular}{c c c}         
  \hline \hline 
Band & Observations & Model  \\
\hline
   3.3  & 2.5$\pm$0.5 & 10 \\
 3.4  &    0.6  $\pm$0.1       & 2.5 \\
   6.2  &  14.3$\pm$0.4           & 37  \\
   8.6  & 4.6$\pm$0.5  & 12 \\
\hline
\end{tabular}
\end{table}

\subsection{Predicted spatial distribution compared to NaCo data}
\label{sec_model_obs_2}

\begin{figure}[htbp]
\resizebox{\hsize}{!}{\includegraphics[width=\textwidth,height=\textheight,keepaspectratio]{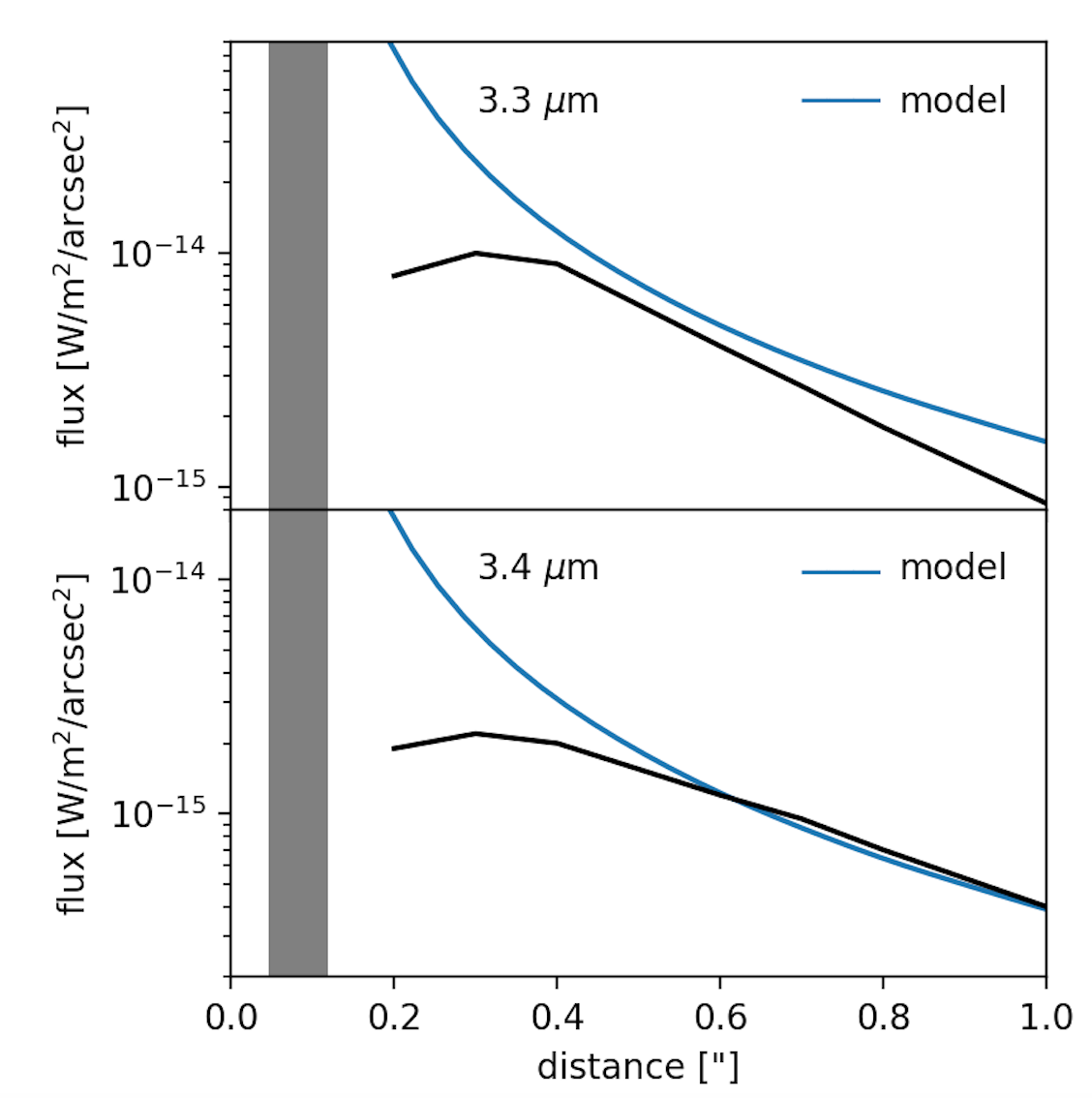}}
\caption{$I_{3.3\mu{\rm m}}$ (top panel) and  $I_{3.4\mu{\rm m}}$ (bottom panel) as observed by NaCo (black) and as predicted by model (blue). The fluxes are the sum of the integrals of the Gaussians related to each feature and averaged for different distances from the star in 0.1" steps. The transparent grey box on the left represents the distance up to which the cavity is extended in the POLARIS simulations: $\sim ~0.12$" or 13~au). The models have been convolved with a 0.1" Gaussian. A mask must be used around the central star due to the high magnitude gradient between on-star spectra and others. 
}
    \label{fig_NaCo_modobs}
\end{figure}

Figure~\ref{fig_NaCo_modobs} presents the predicted and the observed spatial emission profiles of the aromatic and aliphatic bands at 3.3 and 3.4~$\mu$m as a function of the distance to the central star. 
The predicted cubes have been convolved by a Gaussian with a FWHM of 0.1".
The same decomposition with {\tt ROHSA} has been applied to the models to compare them with the observations. To do so, a mask must be used around the central star due to the high magnitude gradient between on-star spectra and spectra farther from the star.

The simulated band emission reproduces roughly the behaviour of the observed profiles, but overestimates the observation of the 3.3~$\mu$m feature at 1" by a factor of 2.
Furthermore, close to the inner rim of the outer disk (between 0.2 and 0.4" or 20 to 40~au), the model does not reproduce the observed plateau and overestimates the intensity by a factor of 2-10. 
In this region, the observed emission is flat and does not vary anymore with $G_0$. This could be due to disk structure effects, such as shadow of the inner edge of the outer disk due to puffed-up inner rim or changes in the properties of the smallest nano-particles, such as a decrease in the a-C1 mass fraction.

\subsection{Predicted spatial distribution compared to VISIR data}
\label{sec_model_obs_3}

\begin{figure}[htbp]
\resizebox{\hsize}{!}{\includegraphics[width=\textwidth,height=\textheight,keepaspectratio]{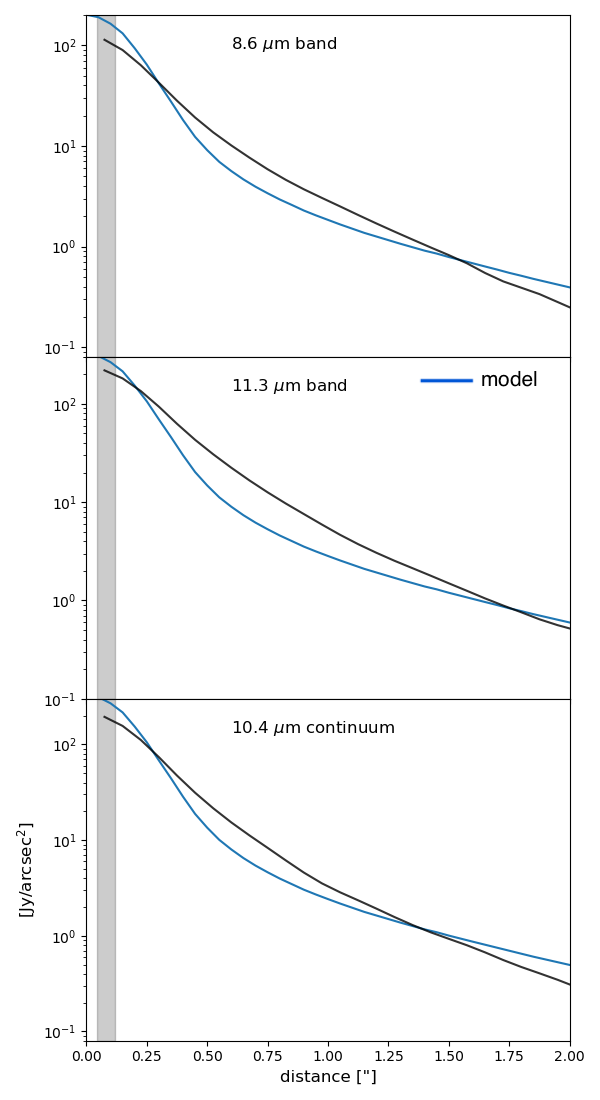}}
    \caption{The 8.6~$\mu$m (top panel), 11.3~$\mu$m (middle panel), and 10.4~$\mu$m (bottom panel) VISIR bands (grey) vs. the model. The predicted profiles have been convolved with a Gaussian of 0.3" similar to the spatial resolution of VISIR and integrated using VISIR filters to compare to the observations. The transparent grey box on the left represents the distance up to which the cavity is extended in the POLARIS simulations: $\sim 0.12$" (13~au).
  }
    \label{fig_visir_modobs}
\end{figure}

Figure~\ref{fig_visir_modobs} presents the predicted and the observed spatial emission profiles in the VISIR bands centred at 8.6 and 11.3~$\mu$m, and in the 10.4~$\mu$m continuum, as a function of the distance to the central star. 
As in the observed emission profiles, the continuum has not been subtracted from the model profiles at 8.6 and 11.3~$\mu$m.
The predicted cubes have been convolved by a Gaussian with a FWHM of 0.3".

In the predicted emission profiles, two regimes appear: thermal emission dominates until 0.5" and then stochastically heated dust emission as $1/d^2$ for $d>0.5$" (50~au), which corresponds to the distance where the model spectra at 10~$\mu$m are mostly due to emission of nano-particles. 
Between 50 and 200~au, the 10 $\mu$m continuum emission is mostly due to the VSG ($ a \sim 10$~nm) and a-C2 ($a \sim 1$~nm) populations, while the 8.6 and 11.3~$\mu$m emission bands are mostly due to a-C1 and a-C2. In term of heating regimes:
\begin{itemize}
\item the a-C1 are stochastically heated;
\item the a-C2 reach thermal equilibrium for $G_0$ of about $10^5$. Then for $G_0 = 10^4-10^5$, they have an intermediate behaviour;
\item the VSG are at thermal equilibrium (whatever the $G_0$).
\end{itemize}

The observed spatial emission profile is fairly well reproduced by our model.
Nevertheless, the model overestimates the flux in the inner regions and in the observed spatial emission profiles, the two emission mechanism regimes appear much less pronounced.  
At large distance, the observed profile does not scale with  $1/d^2$ but appears a bit steeper, as expected if thermal equilibrium emission cannot be neglected. 
Since we did not try to fit the VISIR data with our simulations, these discrepancies could be explained by the parameter choices we made. 
A change in the flux level could be obtained by increasing the mass fraction of the VSG and a-C2 populations for which we have no a priori observational constraints. Indeed, the NaCo data can only constrain the a-C1 population mass fraction (see Sect.~\ref{subsub_rohsa-res}).

\section{Discussion \label{sec_discussion}}

\subsection{Aliphatic-to-aromatic band ratio as a function of the $G_0/n_H$ ratio}

\begin{figure}[htbp]
    \resizebox{\hsize}{!}{\includegraphics[width=\textwidth,height=\textheight,keepaspectratio]{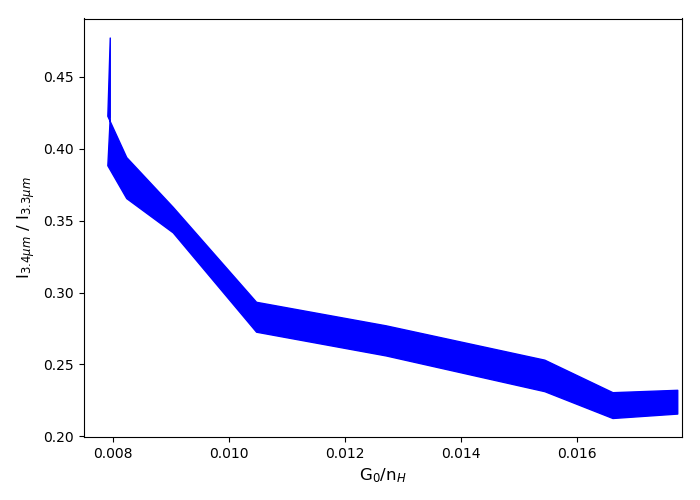}}
    \caption{Spatial profile of the aliphatic-to-aromatic band ratio, $I_{3.4\mu {\rm m}}/I_{3.3\mu {\rm m}}$, according to the $G_0/n_H$ ratio calculated with POLARIS in the disk surface zone from which most of the band emission originates. The highest values of $G_0/n_H$ correspond to the regions closest to the star.}
    \label{fig_aliphatic_aromatic_ratio}
\end{figure}

In Sect.~\ref{sec_obs}, we showed that the aliphatic-to-aromatic band ratio decreases when moving towards the central star. This may reflect the processing of the hydrocarbon grains a-C:H by UV photons. Indeed, laboratory experiments have shown that aliphatic CH bonds are more easily photo-destroyed than aromatic CH bonds \citep[e.g.][]{Munoz2001, Mennella2001, Gadallah2012}. Consequently, a decrease in the aliphatic-to-aromatic band ratio in the inner disk part is expected because the nano-particles are more irradiated.
Nevertheless, we must underline that the aliphatic-to-aromatic band ratio remains quite high, about $\sim 0.2-0.25$, at 0.2" from the central star while the UV field strength is very high, $G_0 \sim 2 \times 10^6$ (see Sect.~\ref{sec_polaris}). According to \citet{Jones2014}, the a-C(:H) photo-processing timescale $\tau_{UV, pd}$ can be analytically expressed as:
\begin{equation}
    \tau_{UV, pd}(a, G_0) = \frac{10^4}{G_0} \left[ 2.7 + \frac{6.5}{(a [{\rm nm}])^{1.4}} + 0.04 a[{\rm nm}]^{1.3}\right] [{\rm yr}],
\end{equation}
which means that the dehydrogenation timescale is expected to be less than about two months for the three populations of a-C(:H) nano-particles.
However, the emission in the 3-4~$\mu$m range depends not only on the UV flux but also on the (re-)hydrogenation rate, which in turn depends on the gas density. Thus, the observation of an aliphatic-to-aromatic band ratio that varies little may suggest a recent exposure of the carriers to the radiation field (by a continuous local replenishment at the disk surface) before their destruction or conversion, and that the $G_0/n_H$ ratio may be a better parameter to consider, instead of the UV field strength only. Figure~\ref{fig_aliphatic_aromatic_ratio} shows the average spatial profile of the aliphatic-to-aromatic band ratio, $I_{3.4\mu {\rm m}}/I_{3.3\mu {\rm m}}$, according to the $G_0/n_H$ calculated with POLARIS in the disk surface zone from which most of the band emission originates (see Fig.~\ref{fig_polaris_g0} in Sect.~\ref{sec_polaris}). 
We find that the aliphatic-to-aromatic band ratio decreases from $\sim 0.45$ to 0.22 for an increase in the $G_0/n_H$ ratio from $\sim 0.008$ to 0.018 as computed from model 1.
This decrease for this range of $G_0/n_H$ values appears consistent with previous studies based on spatially resolved 3-4~$\mu$m spectra towards extended PDRs \citep{Pilleri2015} and other PPDs \citep{Bouteraon2019}. 

In order to interpret more accurately the dependency of the $I_{3.4\mu {\rm m}}/I_{3.3\mu {\rm m}}$ ratio as function of the local physical conditions ($G_0$, $n_H$) a more detailed modelling study is needed. Depending on the photon-grain interaction site and on the photon energy, heating, H atom loss, rehydrogenation and fragmentation should be characterised to determine the time dependent size distribution and composition as a function of $G_0$ and $n_H$. This is however beyond the scope of the present study.

\subsection{The 3-4 $\mu$m spectra near the star and spatial distribution of the 3.52~$\mu$m band \label{ssec_inner}}

This work is mostly focused on the aromatic and aliphatic bands at 3.3 and 3.4~$\mu$m located between 20 and 100~au from the central star. However, it should be noted that the spectra taken in the inner 20~au of the disk are quite different compared to the remaining disk. Indeed, over a strong continuum emission due to hot large grains, the NaCo spectra centred on the star ($r < 20$~au) shows a possible faint aromatic band at 3.3~$\mu$m as well as a faint band at 3.52~$\mu$m. On the other hand, there is no clear evidence of aliphatic bands at 3.4, 3.43, 3.46 and 3.56~$\mu$m. This is in accordance with the analysis of the rest of the disk showing that the $I_{3.4\mu {\rm m}}/I_{3.3\mu {\rm m}}$ ratio decreases when moving closer to the star. 
The detection of a faint band at 3.52~$\mu$m in the spectra centred on the star appears also consistent with the map at 3.52~$\mu$m obtained with {\tt ROHSA} showing a more centrally concentrated emission distribution than the 3.3 and 3.4~$\mu$m features (see Fig.~\ref{fig_rohsa_other-features} in appendix).

The 3.52~$\mu$m band could be emission from nano-diamonds, such as suggested in the HD~97048 and Elias~1 PPD system, where strong emission features at 3.4-3.5 $\mu$m have been detected and are found to be more centrally concentrated than the aromatic emission \citep{Habart2004, Goto2009}. However, higher angular resolution observations probing the carbon dust features in the innermost regions are needed. The  Multi AperTure mid-Infrared SpectroScopic Experiment (MATISSE) instrument offers from 2019 on the possibility to perform interferometric observations with a medium spectral resolution in the 3-4 and 8-13 $\mu$m domain. It will teach us about the spatial distribution and dust properties in the scale of 0.1-10~au.

\section{Conclusions \label{sec_conclusion}}

We have investigated the nano-grain emission in the (pre-)transitional disk HD~100546 by comparing high angular spectroscopic observations with model predictions. 
We analyse adaptive optics spectroscopic data (VLT/NAOS- CONICA, 3-4 $\mu$m, angular resolution 0.1") as well as imaging and spectroscopic data (VLT/VISIR, 8-12 $\mu$m, angular resolution 0.3"). 
The NaCo spectral cube is decomposed using a tool for hyper-spectral analysis ({\tt ROHSA}) in order to produce the first emission maps in both the aromatic and aliphatic features in the 3-4~$\mu$m range.
In our modelling, the THEMIS dust model is integrated into the radiative transfer code POLARIS to calculate the thermal and stochastic heating of micro- and nanometric dust grains inside a disk structure.
In the THEMIS model, the carbon grains with sizes from 0.4 to 20~nm have continuous size-dependent optical properties as well as a continuous size distribution contrary to most dust models. 
This allows us to model the sub- and main features characterising bonds between carbon and hydrogen atoms in the 3-4 $\mu$m range. These features strongly depend on the grain size distribution and the hydrogen-to-carbon ratio in the grains, depending on their more or less aromatic or aliphatic nature. They also vary with the local environment properties: $G_0$ and $n_H$.
 Our main results derived from both observations and model predictions can be summarised as follows.

\begin{enumerate}
    \item The aromatic and aliphatic features between 3.3 and 3.5~$\mu$m are spatially extended with each band showing a unique morphology. 
    
    \item The aliphatic-to-aromatic band ratio, 3.4/3.3, increases with the distance to the star from $\sim$0.2 (at 0.2" or 20~au) to $\sim$0.45 (at 1" or 100~au). This is in agreement with previous studies and argues for an evolution of the composition of nano-materials which are more aromatic in the disk parts close to the star, where the UV field ($G_0$) is high.

    \item The model predicts that the aromatic and aliphatic bands are due to different combinations of grain sub-populations with a high dependence on the intensity of the UV radiation field ($G_0$). The 3.3 and 3.4 $\mu$m features are mainly due the extremely small nano-particles with $a\le$ 0.7~nm both at low and high $G_0$. On the other hand, the 8.6~ and 11.3~$\mu$m features are mainly due to the larger nano-particles with $a\le$1.5~nm at low $G_0$, and a mix of the contribution of several grain populations at high $G_0$.  The 10.4~$\mu$m continuum is mainly 
    due to the sub-micronic grains near the star, while at large distance, the largest nano-particles ($a\ge$1.5 nm) contribute to $\sim 50$\%, and nano-particles with $a\le$1.5 nm and $a\le$ 0.7 nm to $\sim 40$\% and $\sim 10$\% respectively.
    
    \item The observed continuum emission at 10~$\mu$m is spatially extended, as in the bands at 8.6 and 11.3 $\mu$m. This is in agreement with the model, which predicts that at large distance the continuum emission is mainly caused by nano-particles of extremely small to larger sizes.
     
    \item In the observed spectra in the 8-12 $\mu$m range, several aromatic and crystalline silicates features are detected  up to a large distance (at least 1.5"). The emission in the aromatic bands increase between 0.5" and 1.5". The crystalline features at large distance could suggest the presence of nano-silicates.

    \item We investigated a disk model that reproduces well the spatial emission profiles of the 3.3 and 3.4~$\mu$m bands, except for the inner 20-40~au where the observed emission is, unlike the predictions, flat and does not increase anymore with the UV field. 
    The possible explanations are disk structure effects (e.g. shadow of the inner edge of the outer disk) or a decrease in the mass fraction in the nano-particles with $a \le 0.7$~nm mainly responsible for the 3.3 and 3.4~$\mu$m features. The model reproduces fairly well the observed spatial emission profiles in the mid-IR at 8.6, 10.4 and 11.3 $\mu$m. 
\end{enumerate}

Our understanding of the evolution of carbonaceous nano-grains in PPDs is expected to make significant progress by constraints imposed by upcoming spatially resolved spectroscopic observations of the carbon nano-grains in protoplanetary disks with the VLTI/MATISSE and JWST.
Interferometric observations could be used to constrain the spatial distribution and properties of carbonaceous material in the terrestrial planet forming region. 
On the other hand, observations with the JWST covering a complete spectral domain between 0.6 and 28~$\mu$m will mainly probe the warm gas and small dust content with a spatial resolution of 10-100~au and for distances up to 500~au. Spatially resolved spectroscopy of the sub-features from aromatic/PAHs, aliphatics will be obtained and permit to better identify the nature of the band carriers and the main processes. 

\begin{acknowledgements}
This work was based on observations collected at the European Southern Observatory, Chile (ESO proposal number: 075.C-0624(A)), and was supported by P2IO LabEx (ANR-10-LABX-0038) in the framework of the "Investissements d'Avenir" (ANR-11-IDEX-0003-01) managed by the Agence Nationale de la Recherche (ANR, France), P2IO LabEx (A-JWST-01-02-01+LABEXP2IOPDO), and Programme National "Physique et Chimie du Milieu Interstellaire" (PCMI) of CNRS/INSU with INC/INP co-funded by CEA and CNES.
\end{acknowledgements}

\bibliographystyle{aa} 
\bibliography{ref.bib}

\begin{thebibliography}{79}
\expandafter\ifx\csname natexlab\endcsname\relax\def\natexlab#1{#1}\fi

\bibitem[{{Acke} {et~al.}(2010){Acke}, {Bouwman}, {Juh{\'a}sz}, {Henning}, {van
  den Ancker}, {Meeus}, {Tielens}, \& {Waters}}]{Acke2010}
{Acke}, B., {Bouwman}, J., {Juh{\'a}sz}, A., {et~al.} 2010, \apj, 718, 558

\bibitem[{Acke \& van~den Ancker(2004)}]{acke_iso_2004}
Acke, B. \& van~den Ancker, M.~E. 2004, 426, 151

\bibitem[{{Ardila} {et~al.}(2007){Ardila}, {Golimowski}, {Krist}, {Clampin},
  {Ford}, \& {Illingworth}}]{Ardila2007}
{Ardila}, D.~R., {Golimowski}, D.~A., {Krist}, J.~E., {et~al.} 2007, \apj, 665,
  512

\bibitem[{{Augereau} {et~al.}(2001){Augereau}, {Lagrange}, {Mouillet}, \&
  {M{\'e}nard}}]{Augereau2001}
{Augereau}, J.~C., {Lagrange}, A.~M., {Mouillet}, D., \& {M{\'e}nard}, F. 2001,
  \aap, 365, 78

\bibitem[{{Avenhaus} {et~al.}(2014){Avenhaus}, {Quanz}, {Meyer}, {Brittain},
  {Carr}, \& {Najita}}]{Avenhaus2014}
{Avenhaus}, H., {Quanz}, S.~P., {Meyer}, M.~R., {et~al.} 2014, \apj, 790, 56

\bibitem[{{Benisty} {et~al.}(2010){Benisty}, {Tatulli}, {M{\'e}nard}, \&
  {Swain}}]{Benisty2010}
{Benisty}, M., {Tatulli}, E., {M{\'e}nard}, F., \& {Swain}, M.~R. 2010, \aap,
  511, A75

\bibitem[{{Boccaletti} {et~al.}(2013){Boccaletti}, {Pantin}, {Lagrange},
  {Augereau}, {Meheut}, \& {Quanz}}]{Boccaletti2013}
{Boccaletti}, A., {Pantin}, E., {Lagrange}, A.~M., {et~al.} 2013, \aap, 560,
  A20

\bibitem[{{Bocchio} {et~al.}(2014){Bocchio}, {Jones}, \&
  {Slavin}}]{Bocchio2014}
{Bocchio}, M., {Jones}, A.~P., \& {Slavin}, J.~D. 2014, \aap, 570, A32

\bibitem[{{Bout{\'e}raon} {et~al.}(2019){Bout{\'e}raon}, {Habart}, {Ysard},
  {Jones}, {Dartois}, \& {Pino}}]{Bouteraon2019}
{Bout{\'e}raon}, T., {Habart}, E., {Ysard}, N., {et~al.} 2019, \aap, 623, A135

\bibitem[{{Bouwman} {et~al.}(2003){Bouwman}, {de Koter}, {Dominik}, \&
  {Waters}}]{Bouwman2003}
{Bouwman}, J., {de Koter}, A., {Dominik}, C., \& {Waters}, L.~B.~F.~M. 2003,
  \aap, 401, 577

\bibitem[{{Brauer} {et~al.}(2019){Brauer}, {Pantin}, {Di Folco}, {Habart},
  {Dutrey}, \& {Guilloteau}}]{Brauer2019}
{Brauer}, R., {Pantin}, E., {Di Folco}, E., {et~al.} 2019, \aap, 628, A88

\bibitem[{{Brauer} {et~al.}(2017){Brauer}, {Wolf}, {Reissl}, \&
  {Ober}}]{brauer_magnetic_2017-1}
{Brauer}, R., {Wolf}, S., {Reissl}, S., \& {Ober}, F. 2017, \aap, 601, A90

\bibitem[{{Brittain} {et~al.}(2009){Brittain}, {Najita}, \&
  {Carr}}]{Brittain2009}
{Brittain}, S.~D., {Najita}, J.~R., \& {Carr}, J.~S. 2009, \apj, 702, 85

\bibitem[{{Brooke} {et~al.}(1993){Brooke}, {Tokunaga}, \& {Strom}}]{Brooke1993}
{Brooke}, T.~Y., {Tokunaga}, A.~T., \& {Strom}, S.~E. 1993, \aj, 106, 656

\bibitem[{{Carmona} {et~al.}(2011){Carmona}, {van der Plas}, {van den Ancker},
  {Audard}, {Waters}, {Fedele}, {Acke}, \& {Pantin}}]{Carmona2011}
{Carmona}, A., {van der Plas}, G., {van den Ancker}, M.~E., {et~al.} 2011,
  \aap, 533, A39

\bibitem[{{Chastenet} {et~al.}(2017){Chastenet}, {Bot}, {Gordon}, {Bocchio},
  {Roman-Duval}, {Jones}, \& {Ysard}}]{Chastenet2017}
{Chastenet}, J., {Bot}, C., {Gordon}, K.~D., {et~al.} 2017, \aap, 601, A55

\bibitem[{{Chiang} \& {Goldreich}(1997)}]{chiang1997}
{Chiang}, E.~I. \& {Goldreich}, P. 1997, \apj, 490, 368

\bibitem[{{Cohen} {et~al.}(1999){Cohen}, {Walker}, {Carter}, {Hammersley},
  {Kidger}, \& {Noguchi}}]{Cohen1999}
{Cohen}, M., {Walker}, R.~G., {Carter}, B., {et~al.} 1999, \aj, 117, 1864

\bibitem[{{Compi{\`e}gne} {et~al.}(2011){Compi{\`e}gne}, {Verstraete}, {Jones},
  {Bernard}, {Boulanger}, {Flagey}, {Le Bourlot}, {Paradis}, \&
  {Ysard}}]{Compiegne2011}
{Compi{\`e}gne}, M., {Verstraete}, L., {Jones}, A., {et~al.} 2011, \aap, 525,
  A103

\bibitem[{{Croiset} {et~al.}(2016){Croiset}, {Candian}, {Bern{\'e}}, \&
  {Tielens}}]{croiset2016}
{Croiset}, B.~A., {Candian}, A., {Bern{\'e}}, O., \& {Tielens}, A.~G.~G.~M.
  2016, \aap, 590, A26

\bibitem[{{Currie} {et~al.}(2015){Currie}, {Cloutier}, {Brittain}, {Grady},
  {Burrows}, {Muto}, {Kenyon}, \& {Kuchner}}]{Currie2015}
{Currie}, T., {Cloutier}, R., {Brittain}, S., {et~al.} 2015, \apjl, 814, L27

\bibitem[{{Desert} {et~al.}(1990){Desert}, {Boulanger}, \&
  {Puget}}]{Desert1990}
{Desert}, F.~X., {Boulanger}, F., \& {Puget}, J.~L. 1990, \aap, 500, 313

\bibitem[{Draine \& Li(2001)}]{draine_infrared_2001}
Draine, B.~T. \& Li, A. 2001, The Astrophysical Journal, 551, 807

\bibitem[{Draine \& Li(2007)}]{draine_infrared_2007}
Draine, B.~T. \& Li, A. 2007, The Astrophysical Journal, 657, 810

\bibitem[{{Follette} {et~al.}(2017){Follette}, {Rameau}, {Dong}, {Pueyo},
  {Close}, {Duch{\^e}ne}, {Fung}, {Leonard}, {Macintosh}, {Males}, {Marois},
  {Millar-Blanchaer}, {Morzinski}, {Mullen}, {Perrin}, {Spiro}, {Wang},
  {Ammons}, {Bailey}, {Barman}, {Bulger}, {Chilcote}, {Cotten}, {De Rosa},
  {Doyon}, {Fitzgerald}, {Goodsell}, {Graham}, {Greenbaum}, {Hibon}, {Hung},
  {Ingraham}, {Kalas}, {Konopacky}, {Larkin}, {Maire}, {Marchis}, {Metchev},
  {Nielsen}, {Oppenheimer}, {Palmer}, {Patience}, {Poyneer}, {Rajan},
  {Rantakyr{\"o}}, {Savransky}, {Schneider}, {Sivaramakrishnan}, {Song},
  {Soummer}, {Thomas}, {Vega}, {Wallace}, {Ward-Duong}, {Wiktorowicz}, \&
  {Wolff}}]{Follette2017}
{Follette}, K.~B., {Rameau}, J., {Dong}, R., {et~al.} 2017, \aj, 153, 264

\bibitem[{{Gadallah} {et~al.}(2012){Gadallah}, {Mutschke}, \&
  {J{\"a}ger}}]{Gadallah2012}
{Gadallah}, K.~A.~K., {Mutschke}, H., \& {J{\"a}ger}, C. 2012, \aap, 544, A107

\bibitem[{{Gaia Collaboration} {et~al.}(2018){Gaia Collaboration}, {Brown},
  {Vallenari}, {Prusti}, {de Bruijne}, {Babusiaux}, {Bailer-Jones}, {Biermann},
  {Evans}, {Eyer}, \& et~al.}]{Gaia2018}
{Gaia Collaboration}, {Brown}, A.~G.~A., {Vallenari}, A., {et~al.} 2018, \aap,
  616, A1

\bibitem[{{Garufi} {et~al.}(2016){Garufi}, {Quanz}, {Schmid}, {Mulders},
  {Avenhaus}, {Boccaletti}, {Ginski}, {Langlois}, {Stolker}, {Augereau},
  {Benisty}, {Lopez}, {Dominik}, {Gratton}, {Henning}, {Janson}, {M{\'e}nard},
  {Meyer}, {Pinte}, {Sissa}, {Vigan}, {Zurlo}, {Bazzon}, {Buenzli}, {Bonnefoy},
  {Brandner}, {Chauvin}, {Cheetham}, {Cudel}, {Desidera}, {Feldt}, {Galicher},
  {Kasper}, {Lagrange}, {Lannier}, {Maire}, {Mesa}, {Mouillet}, {Peretti},
  {Perrot}, {Salter}, \& {Wildi}}]{Garufi2016}
{Garufi}, A., {Quanz}, S.~P., {Schmid}, H.~M., {et~al.} 2016, \aap, 588, A8

\bibitem[{{Geers} {et~al.}(2007){Geers}, {Pontoppidan}, {van Dishoeck},
  {Dullemond}, {Augereau}, {Mer{\'\i}n}, {Oliveira}, \& {Pel}}]{Geers2007}
{Geers}, V.~C., {Pontoppidan}, K.~M., {van Dishoeck}, E.~F., {et~al.} 2007,
  \aap, 469, L35

\bibitem[{{Gorti} \& {Hollenbach}(2008)}]{gorti2008}
{Gorti}, U. \& {Hollenbach}, D. 2008, \apj, 683, 287

\bibitem[{{Goto} {et~al.}(2009){Goto}, {Henning}, {Kouchi}, {Takami}, {Hayano},
  {Usuda}, {Takato}, {Terada}, {Oya}, {J{\"a}ger}, \& {Andersen}}]{Goto2009}
{Goto}, M., {Henning}, T., {Kouchi}, A., {et~al.} 2009, \apj, 693, 610

\bibitem[{{Grady} {et~al.}(2001){Grady}, {Polomski}, {Henning}, {Stecklum},
  {Woodgate}, {Telesco}, {Pi{\~n}a}, {Gull}, {Boggess}, {Bowers}, {Bruhweiler},
  {Clampin}, {Danks}, {Green}, {Heap}, {Hutchings}, {Jenkins}, {Joseph},
  {Kaiser}, {Kimble}, {Kraemer}, {Lindler}, {Linsky}, {Maran}, {Moos}, {Plait},
  {Roesler}, {Timothy}, \& {Weistrop}}]{Grady2001}
{Grady}, C.~A., {Polomski}, E.~F., {Henning}, T., {et~al.} 2001, \aj, 122, 3396

\bibitem[{{Grady} {et~al.}(2005){Grady}, {Woodgate}, {Heap}, {Bowers}, {Nuth},
  {Herczeg}, \& {Hill}}]{Grady2005}
{Grady}, C.~A., {Woodgate}, B., {Heap}, S.~R., {et~al.} 2005, \apj, 620, 470

\bibitem[{{Habart} {et~al.}(2004){Habart}, {Natta}, \&
  {Kr{\"u}gel}}]{Habart2004}
{Habart}, E., {Natta}, A., \& {Kr{\"u}gel}, E. 2004, \aap, 427, 179

\bibitem[{{Habart} {et~al.}(2006){Habart}, {Natta}, {Testi}, \&
  {Carbillet}}]{Habart2006}
{Habart}, E., {Natta}, A., {Testi}, L., \& {Carbillet}, M. 2006, \aap, 449,
  1067

\bibitem[{{Habing}(1968)}]{Habing1968}
{Habing}, H.~J. 1968, \bain, 19, 421

\bibitem[{Hayashi(1981)}]{hayashi_structure_1981}
Hayashi, C. 1981, Progress of Theoretical Physics Supplement, 70, 35

\bibitem[{{Hu} {et~al.}(1989){Hu}, {The}, \& {de Winter}}]{Hu1989}
{Hu}, J.~Y., {The}, P.~S., \& {de Winter}, D. 1989, \aap, 208, 213

\bibitem[{{Jones} {et~al.}(2013){Jones}, {Fanciullo}, {K{\"o}hler},
  {Verstraete}, {Guillet}, {Bocchio}, \& {Ysard}}]{Jones2013}
{Jones}, A.~P., {Fanciullo}, L., {K{\"o}hler}, M., {et~al.} 2013, \aap, 558,
  A62

\bibitem[{{Jones} {et~al.}(2017){Jones}, {K{\"o}hler}, {Ysard}, {Bocchio}, \&
  {Verstraete}}]{Jones2017}
{Jones}, A.~P., {K{\"o}hler}, M., {Ysard}, N., {Bocchio}, M., \& {Verstraete},
  L. 2017, \aap, 602, A46

\bibitem[{{Jones} {et~al.}(2014){Jones}, {Ysard}, {K{\"o}hler}, {Fanciullo},
  {Bocchio}, {Micelotta}, {Verstraete}, \& {Guillet}}]{Jones2014}
{Jones}, A.~P., {Ysard}, N., {K{\"o}hler}, M., {et~al.} 2014, Faraday
  Discussions, 168, 313

\bibitem[{{Klarmann} {et~al.}(2017){Klarmann}, {Benisty}, {Min}, {Dominik},
  {Berger}, {Waters}, {Kluska}, {Lazareff}, \& {Le Bouquin}}]{Klarmann2017}
{Klarmann}, L., {Benisty}, M., {Min}, M., {et~al.} 2017, \aap, 599, A80

\bibitem[{{Kluska} {et~al.}(2018){Kluska}, {Kraus}, {Davies}, {Harries},
  {Willson}, {Monnier}, {Aarnio}, {Baron}, {Millan-Gabet}, {Ten Brummelaar},
  {Che}, {Hinkley}, {Preibisch}, {Sturmann}, {Sturmann}, \&
  {Touhami}}]{Kluska2018}
{Kluska}, J., {Kraus}, S., {Davies}, C.~L., {et~al.} 2018, \apj, 855, 44

\bibitem[{{K{\"o}hler} {et~al.}(2014){K{\"o}hler}, {Jones}, \&
  {Ysard}}]{Koehler2014}
{K{\"o}hler}, M., {Jones}, A., \& {Ysard}, N. 2014, \aap, 565, L9

\bibitem[{{K{\"o}hler} {et~al.}(2015){K{\"o}hler}, {Ysard}, \&
  {Jones}}]{Koehler2015}
{K{\"o}hler}, M., {Ysard}, N., \& {Jones}, A.~P. 2015, \aap, 579, A15

\bibitem[{{Kraus} {et~al.}(2013){Kraus}, {Ireland}, {Sitko}, {Monnier},
  {Calvet}, {Espaillat}, {Grady}, {Harries}, {H{\"o}nig}, {Russell},
  {Swearingen}, {Werren}, \& {Wilner}}]{Kraus2013}
{Kraus}, S., {Ireland}, M.~J., {Sitko}, M.~L., {et~al.} 2013, \apj, 768, 80

\bibitem[{{Lagage} {et~al.}(2006){Lagage}, {Doucet}, {Pantin}, {Habart},
  {Duch{\^e}ne}, {M{\'e}nard}, {Pinte}, {Charnoz}, \& {Pel}}]{Lagage2006}
{Lagage}, P.-O., {Doucet}, C., {Pantin}, E., {et~al.} 2006, Science, 314, 621

\bibitem[{{Li} \& {Draine}(2001)}]{Li2001}
{Li}, A. \& {Draine}, B.~T. 2001, \apj, 554, 778

\bibitem[{{Liu} {et~al.}(2003){Liu}, {Hinz}, {Meyer}, {Mamajek}, {Hoffmann}, \&
  {Hora}}]{Liu2003}
{Liu}, W.~M., {Hinz}, P.~M., {Meyer}, M.~R., {et~al.} 2003, \apjl, 598, L111

\bibitem[{{Maaskant} {et~al.}(2013){Maaskant}, {Honda}, {Waters}, {Tielens},
  {Dominik}, {Min}, {Verhoeff}, {Meeus}, \& {van den Ancker}}]{Maaskant2013}
{Maaskant}, K.~M., {Honda}, M., {Waters}, L.~B.~F.~M., {et~al.} 2013, \aap,
  555, A64

\bibitem[{{Malfait} {et~al.}(1998){Malfait}, {Waelkens}, {Waters}, {Vand
  enbussche}, {Huygen}, \& {de Graauw}}]{Malfait1998}
{Malfait}, K., {Waelkens}, C., {Waters}, L.~B.~F.~M., {et~al.} 1998, \aap, 332,
  L25

\bibitem[{{Marchal} {et~al.}(2019){Marchal}, {Miville-Desch{\^e}nes}, {Orieux},
  {Gac}, {Soussen}, {Lesot}, {d'Allonnes}, \& {Salom{\'e}}}]{marchal2019}
{Marchal}, A., {Miville-Desch{\^e}nes}, M.-A., {Orieux}, F., {et~al.} 2019,
  \aap, 626, A101

\bibitem[{{Meeus} {et~al.}(2012){Meeus}, {Montesinos}, {Mendigut{\'\i}a},
  {Kamp}, {Thi}, {Eiroa}, {Grady}, {Mathews}, {Sandell}, {Martin-Za{\"\i}di},
  {Brittain}, {Dent}, {Howard}, {M{\'e}nard}, {Pinte}, {Roberge}, {Vand
  enbussche}, \& {Williams}}]{Meeus2012}
{Meeus}, G., {Montesinos}, B., {Mendigut{\'\i}a}, I., {et~al.} 2012, \aap, 544,
  A78

\bibitem[{{Meeus} {et~al.}(2013){Meeus}, {Salyk}, {Bruderer}, {Fedele},
  {Maaskant}, {Evans}, {van Dishoeck}, {Montesinos}, {Herczeg}, {Bouwman},
  {Green}, {Dominik}, {Henning}, \& {Vicente}}]{Meeus2013}
{Meeus}, G., {Salyk}, C., {Bruderer}, S., {et~al.} 2013, \aap, 559, A84

\bibitem[{{Meeus} {et~al.}(2001){Meeus}, {Waters}, {Bouwman}, {van den Ancker},
  {Waelkens}, \& {Malfait}}]{Meeus2001}
{Meeus}, G., {Waters}, L.~B.~F.~M., {Bouwman}, J., {et~al.} 2001, \aap, 365,
  476

\bibitem[{{Mennella} {et~al.}(2001){Mennella}, {Mu{\~n}oz Caro}, {Ruiterkamp},
  {Schutte}, {Greenberg}, {Brucato}, \& {Colangeli}}]{Mennella2001}
{Mennella}, V., {Mu{\~n}oz Caro}, G.~M., {Ruiterkamp}, R., {et~al.} 2001, \aap,
  367, 355

\bibitem[{{Miley} {et~al.}(2019){Miley}, {Pani{\'c}}, {Haworth}, {Pascucci},
  {Wyatt}, {Clarke}, {Richards}, \& {Ratzka}}]{Miley2019}
{Miley}, J.~M., {Pani{\'c}}, O., {Haworth}, T.~J., {et~al.} 2019, \mnras, 485,
  739

\bibitem[{{Mu{\~n}oz Caro} {et~al.}(2001){Mu{\~n}oz Caro}, {Ruiterkamp},
  {Schutte}, {Greenberg}, \& {Mennella}}]{Munoz2001}
{Mu{\~n}oz Caro}, G.~M., {Ruiterkamp}, R., {Schutte}, W.~A., {Greenberg},
  J.~M., \& {Mennella}, V. 2001, \aap, 367, 347

\bibitem[{{Mulders} {et~al.}(2013){Mulders}, {Paardekooper}, {Pani{\'c}},
  {Dominik}, {van Boekel}, \& {Ratzka}}]{Mulders2013}
{Mulders}, G.~D., {Paardekooper}, S.-J., {Pani{\'c}}, O., {et~al.} 2013, \aap,
  557, A68

\bibitem[{{Pani{\'c}} {et~al.}(2014){Pani{\'c}}, {Ratzka}, {Mulders},
  {Dominik}, {van Boekel}, {Henning}, {Jaffe}, \& {Min}}]{Panic2014}
{Pani{\'c}}, O., {Ratzka}, T., {Mulders}, G.~D., {et~al.} 2014, \aap, 562, A101

\bibitem[{{Pantin} {et~al.}(2000){Pantin}, {Waelkens}, \&
  {Lagage}}]{Pantin2000}
{Pantin}, E., {Waelkens}, C., \& {Lagage}, P.~O. 2000, \aap, 361, L9

\bibitem[{{P{\'e}rez} {et~al.}(2019){P{\'e}rez}, {Casassus}, {Hales}, {Marino},
  {Cheetham}, {Zurlo}, {Cieza}, {Dong}, {Alarc{\'o}n}, {Ben{\'\i}tez-Llambay},
  \& {Fomalont}}]{Perez2019}
{P{\'e}rez}, S., {Casassus}, S., {Hales}, A., {et~al.} 2019, arXiv e-prints,
  arXiv:1906.06305

\bibitem[{{Pilleri} {et~al.}(2015){Pilleri}, {Joblin}, {Boulanger}, \&
  {Onaka}}]{Pilleri2015}
{Pilleri}, P., {Joblin}, C., {Boulanger}, F., \& {Onaka}, T. 2015, \aap, 577,
  A16

\bibitem[{{Pineda} {et~al.}(2019){Pineda}, {Szul{\'a}gyi}, {Quanz}, {van
  Dishoeck}, {Garufi}, {Meru}, {Mulders}, {Testi}, {Meyer}, \&
  {Reggiani}}]{Pineda2019}
{Pineda}, J.~E., {Szul{\'a}gyi}, J., {Quanz}, S.~P., {et~al.} 2019, \apj, 871,
  48

\bibitem[{{Quanz} {et~al.}(2015){Quanz}, {Amara}, {Meyer}, {Girard},
  {Kenworthy}, \& {Kasper}}]{Quanz2015}
{Quanz}, S.~P., {Amara}, A., {Meyer}, M.~R., {et~al.} 2015, \apj, 807, 64

\bibitem[{{Rameau} {et~al.}(2017){Rameau}, {Follette}, {Pueyo}, {Marois},
  {Macintosh}, {Millar-Blanchaer}, {Wang}, {Vega}, {Doyon}, {Lafreni{\`e}re},
  {Nielsen}, {Bailey}, {Chilcote}, {Close}, {Esposito}, {Males}, {Metchev},
  {Morzinski}, {Ruffio}, {Wolff}, {Ammons}, {Barman}, {Bulger}, {Cotten}, {De
  Rosa}, {Duchene}, {Fitzgerald}, {Goodsell}, {Graham}, {Greenbaum}, {Hibon},
  {Hung}, {Ingraham}, {Kalas}, {Konopacky}, {Larkin}, {Maire}, {Marchis},
  {Oppenheimer}, {Palmer}, {Patience}, {Perrin}, {Poyneer}, {Rajan},
  {Rantakyr{\"o}}, {Marley}, {Savransky}, {Schneider}, {Sivaramakrishnan},
  {Song}, {Soummer}, {Thomas}, {Wallace}, {Ward-Duong}, \&
  {Wiktorowicz}}]{Rameau2017}
{Rameau}, J., {Follette}, K.~B., {Pueyo}, L., {et~al.} 2017, \aj, 153, 244

\bibitem[{{Reissl} {et~al.}(2016){Reissl}, {Wolf}, \& {Brauer}}]{Reissl2016}
{Reissl}, S., {Wolf}, S., \& {Brauer}, R. 2016, \aap, 593, A87

\bibitem[{{Seok} \& {Li}(2017)}]{Seok2017}
{Seok}, J.~Y. \& {Li}, A. 2017, \apj, 835, 291

\bibitem[{{Shakura} \& {Sunyaev}(1973)}]{Shakura1973}
{Shakura}, N.~I. \& {Sunyaev}, R.~A. 1973, \aap, 500, 33

\bibitem[{{Sturm} {et~al.}(2010){Sturm}, {Bouwman}, {Henning}, {Evans}, {Acke},
  {Mulders}, {Waters}, {van Dishoeck}, {Meeus}, {Green}, {Augereau},
  {Olofsson}, {Salyk}, {Najita}, {Herczeg}, {van Kempen}, {Kristensen},
  {Dominik}, {Carr}, {Waelkens}, {Bergin}, {Blake}, {Brown}, {Chen}, {Cieza},
  {Dunham}, {Glassgold}, {G{\"u}del}, {Harvey}, {Hogerheijde}, {Jaffe},
  {J{\o}rgensen}, {Kim}, {Knez}, {Lacy}, {Lee}, {Maret}, {Meijerink},
  {Mer{\'\i}n}, {Mundy}, {Pontoppidan}, {Visser}, \&
  {Y{\i}ld{\i}z}}]{sturm2010}
{Sturm}, B., {Bouwman}, J., {Henning}, T., {et~al.} 2010, \aap, 518, L129

\bibitem[{{Tatulli} {et~al.}(2011){Tatulli}, {Benisty}, {M{\'e}nard},
  {Varni{\`e}re}, {Martin-Za{\"\i}di}, {Thi}, {Pinte}, {Massi}, {Weigelt},
  {Hofmann}, \& {Petrov}}]{Tatulli2011}
{Tatulli}, E., {Benisty}, M., {M{\'e}nard}, F., {et~al.} 2011, \aap, 531, A1

\bibitem[{{Thi} {et~al.}(2011){Thi}, {M{\'e}nard}, {Meeus},
  {Martin-Za{\"\i}di}, {Woitke}, {Tatulli}, {Benisty}, {Kamp}, {Pascucci},
  {Pinte}, {Grady}, {Brittain}, {White}, {Howard}, {Sandell}, \&
  {Eiroa}}]{Thi2011}
{Thi}, W.~F., {M{\'e}nard}, F., {Meeus}, G., {et~al.} 2011, \aap, 530, L2

\bibitem[{{van Boekel} {et~al.}(2004){van Boekel}, {Waters}, {Dominik},
  {Dullemond}, {Tielens}, \& {de Koter}}]{VanBoekel2004}
{van Boekel}, R., {Waters}, L.~B.~F.~M., {Dominik}, C., {et~al.} 2004, \aap,
  418, 177

\bibitem[{{van Boekel}(2004)}]{Boekel2004PhD}
{van Boekel}, R.~J.~H.~M. 2004, PhD thesis, University of Amsterdam

\bibitem[{{van der Plas} {et~al.}(2009){van der Plas}, {van den Ancker},
  {Acke}, {Carmona}, {Dominik}, {Fedele}, \& {Waters}}]{VanDerPlas2009}
{van der Plas}, G., {van den Ancker}, M.~E., {Acke}, B., {et~al.} 2009, \aap,
  500, 1137

\bibitem[{{Van Kerckhoven}(2002)}]{Kerckhoven2002phd}
{Van Kerckhoven}, C. 2002, PhD thesis, Institute of Astronomy, Katholieke
  Universiteit Leuven, Belgium

\bibitem[{{Viaene} {et~al.}(2019){Viaene}, {Sarzi}, {Zabel}, {Coccato},
  {Corsini}, {Davis}, {De Vis}, {de Zeeuw}, {Falc{\'o}n-Barroso}, {Gadotti},
  {Iodice}, {Lyubenova}, {McDermid}, {Morelli}, {Nedelchev}, {Pinna},
  {Spriggs}, \& {van de Ven}}]{viaene2019}
{Viaene}, S., {Sarzi}, M., {Zabel}, N., {et~al.} 2019, \aap, 622, A89

\bibitem[{{Ysard} {et~al.}(2016){Ysard}, {K{\"o}hler}, {Jones}, {Dartois},
  {Godard}, \& {Gavilan}}]{Ysard2016}
{Ysard}, N., {K{\"o}hler}, M., {Jones}, A., {et~al.} 2016, \aap, 588, A44

\bibitem[{{Ysard} {et~al.}(2015){Ysard}, {K{\"o}hler}, {Jones},
  {Miville-Desch{\^e}nes}, {Abergel}, \& {Fanciullo}}]{Ysard2015}
{Ysard}, N., {K{\"o}hler}, M., {Jones}, A., {et~al.} 2015, \aap, 577, A110

\end{thebibliography}

\begin{appendix}

\section{{\tt ROHSA} decomposition \label{app_rohsa}}

\begin{figure*}[htbp]
    \centering
    \includegraphics[width=17cm]{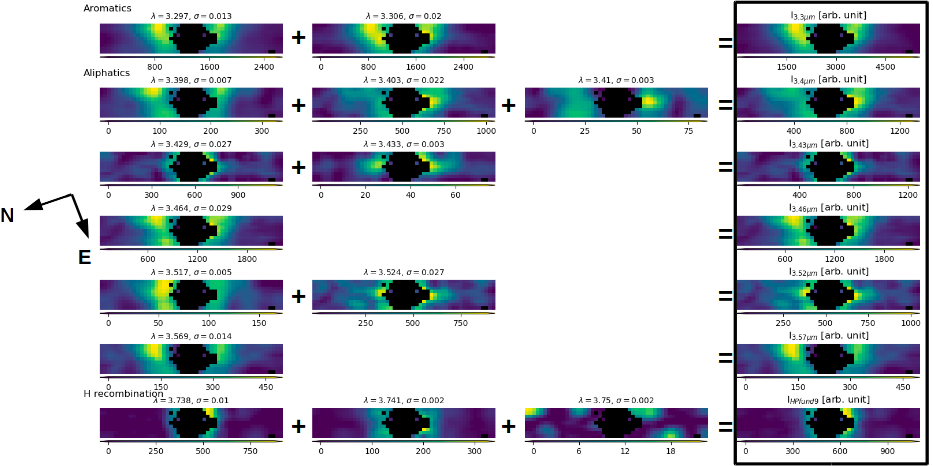}
    \caption{Main feature maps from the ROHSA decomposition of NaCo data. The last column is the sum of the first ones. As seen in Fig.~\ref{fig_rohsa_maps}, several Gaussians are sometimes needed to reproduce the features.}
    \label{fig_rohsa_other-features}
\end{figure*}

Figure~\ref{fig_rohsa_other-features} shows Gaussians related to the main features in the NaCo observations. They are gathered according to their central wavelength and summed up (last column). Similarly to \citet{Bouteraon2019}, we consider six features related to carbonaceous materials at 3.3, 3.4, 3.43, 3.46, 3.52, and 3.56~$\mu$m. Each feature presents a spatial structured distribution.

The 3.52~$\mu$m band peaks close to the central regions. As we discuss in Sect.~\ref{ssec_inner}, this feature is seen close to the star. The 3.43~$\mu$m band distribution is quite similar to the 3.52~$\mu$m one. These two features are usually attributed to nano-diamonds \citep{Goto2009}.

To note, around 3.74~$\mu$m, there is the hydrogen recombination line Pfund 8 which corresponds to a Lorentzian mechanism. In this decomposition, we only use Gaussian functions. That is why two Gaussians are needed to reproduce the signature.

\end{appendix}

\end{document}